\documentclass[article,showpacs,amsmath,amssymb,superscriptaddress]{revtex4}
\input{psfig.sty}     
\begin{document}

\title{Anderson localization in strongly coupled disordered
electron-phonon systems}

\author{Franz X.~Bronold}
\affiliation{Institut f\"ur Physik,
Ernst-Moritz-Arndt-Universit\"at Greifswald,
D-17487 Greifswald,  Germany}
\author{Andreas Alvermann}
\affiliation{Physikalisches Institut,
Universit\"at Bayreuth, 
D-95440 Bayreuth, Germany}
\author{Holger Fehske}
\affiliation{Institut f\"ur Physik,
Ernst-Moritz-Arndt-Universit\"at Greifswald, 
D-17487 Greifswald,  Germany}
\date{\today}  

\begin{abstract}
{ 
Based on the statistical dynamical mean field 
theory, we investigate, in a generic model
for a strongly coupled disordered electron-phonon 
system, the 
competition between polaron formation  
and Anderson localization.  
The statistical dynamical mean field 
approximation maps the lattice problem to an 
ensemble of self-consistently embedded impurity 
problems. It is a probabilistic approach, focusing
on the distribution instead of the average 
values for observables of interest. We solve the
self-consistent equations of the theory with a
Monte-Carlo sampling technique, representing
distributions for random variables by random samples, 
and discuss various ways to determine mobility edges
from the random sample for the local Green function. 
Specifically, we give, as a function of the 
`polaron parameters', such as adiabaticity and 
electron-phonon coupling constants, a detailed discussion
of the localization properties of a single polaron,
using a bare electron as a reference system.}
\end{abstract}


\maketitle

\section{Introduction}

There is a large variety of materials in which due to 
strong electron-phonon coupling electrons and phonons 
lose their identity and form new composite entities:
polarons. Examples of current 
interest are, among others, the high-temperature 
superconducting 
perovskites~(Bar-Yam {\it et al.} 1992, Salje {\it et al.} 1995),
the non-metallic
nickelates and 
bismuthates~(Cheong {\it et al.} 1994, Oyanagi {\it et al.}
2001),
the colossal magneto-resistance 
manganites~(Jin {\it et al.} 1994, Tokura {\it et al.} 1999),
and semiconducting molecular 
crystals~(Parris {\it et al.} 2001).
In all these materials, electronic properties, for 
instance the electronic spectral function or the 
optical conductivity, show substantial 
polaronic effects. The most important signature in
this respect is   
thermally activated transport (along at least one 
of the crystallographic axes).

The polaron concept has been introduced  
by Landau~(1933)
and  ever since played a central 
role in the analysis of strongly coupled electron-phonon
systems. In a deformable lattice, the coupling between 
an electron and the lattice leads to a lattice deformation 
whose potential tends to bind the electron. This process
is called self-trapping, because the potential depends on
the state of the
electron~(Firsov 1975, Rashba 1982, Wellein and Fehske 1998).
Translational 
symmetry is not broken in this process, transport at low 
enough temperatures is still coherent and band-like, 
with a much
larger mass, however, because of the lattice distortion the
electrons have to carry along.  
The self-trapping process, i.e. the formation
of polarons, is therefore only the crossover from a weakly
dressed electronic quasi-particle to a heavily dressed
polaronic quasi-particle. Even the small polaron,
although much less mobile than the bare electron,
is still itinerant and not localized.

Details of the actual materials notwithstanding, the  
Holstein model~(Holstein 1959),
which describes electrons locally 
coupled to dispersionless phonons, captures the essence
of the self-trapping process. Most of the theoretical 
work is therefore directly based on the Holstein 
model~(DeRaedt and Lagendijk 1983, Ranninger and Thibblin 1992, 
Marsiglio 1993, Alexandrov {\it et al.} 1994, Stephan 1996, 
Wellein {\it et al.} 1996, Capone {\it et al.} 97, 
Bonc\u{a} {\it et al.} 1999).
Extensions to long-range electron-phonon couplings on a 
discrete lattice were discussed by Alexandrov and Kornilovitch (1999)
and by Fehske {\it et al.} (2000).
Despite
the extensive numerical 
simulations, 
the properties of 
the Holstein model are not yet fully understood, especially
at finite densities, where the situation changes 
because of the interaction between 
polarons. But even in the extreme dilute limit, where polarons 
in first approximation do not interact, many questions remain.
For instance, little is quantitatively known about polaron 
formation in a disordered environment. 

On the other hand, polaronic materials are complex materials, 
where chemical as well as crystallographic imperfections 
can be quite substantial. Accordingly, the self-trapping
process, producing a heavy quasi-particle, occurs 
in an environment where, strictly speaking, translational symmetry
is broken, and where defects most probably act as nucleation sites
for the formation of polarons. Therefore, a complete theory of polaron
formation has to take the disordered environment into account. 
In particular, the notoriously difficult question of whether polaron
states are band states or localized defect
states can be only meaningfully addressed in a
context which takes the most important effect due to 
disorder, the possibility of 
Anderson localization, explicitly into account.

In his seminal work Anderson (1958)
has shown
that disorder, if sufficiently strong, dramatically changes
the properties of the electronic states in a solid. 
Whereas for small disorder 
the electronic states are extended Bloch waves, for large
disorder electronic states are localized defect states.  
Anderson localization, i.e., the transition from extended to 
localized states, is well understood for noninteracting 
electrons~
[see, for instance, the review articles by Thouless (1974), 
Brezini and Zerki~(1992) and 
Kramer and MacKinnon~(1993)].
It depends strongly on 
dimensionality. In one dimension [and arguably in 
two dimensions~(Abrahams {\it et al.} 1979)]
localized and extended states cannot 
coexist, because infinitesimally small disorder suffices 
to localize all states at once, whereas in three 
dimensions mobility edges separate
localized from extended states. Since the mobility 
edge is of utmost importance for the transport 
properties~(Mott 1968a, Mott 1968b, Mott 1976, Mott 1981),
an enormous amount of experimental 
and theoretical effort has been directed towards a precise
determination of its position.

That Anderson localization might affect polaron formation, and 
vice versa, has been emphasized several times, starting with
Anderson~(1972)
himself, who pointed out that 
the mobility edge might be surrounded by self-trapped states. 
Not much quantitative work exists however to determine,
for instance, the position of the mobility edge and its 
consequences for the conductivity~(Girvin and Jonson 1980, 
Mueller and Thomas 1983),
or the 
character of the states near the mobility 
edge~(Cohen {\it et al.} 1983)
taking electron-phonon coupling 
explicitly into account. Despite some occasionally fruitful 
adaptations of techniques from nonlinear 
dynamics~(Kopidakis {\it et al.} 1996),
where polaron formation is 
considered as an intrinsic interaction-driven 
nonlinearity, the precise mechanisms of the interplay 
of Anderson localization and polaron formation are 
essentially unknown.

As a first step to address this problem with 
techniques developed in the field of strongly
correlated electron systems, Bronold {\it et al.}~(2001) 
applied the dynamical mean 
field theory (DMFT)~(Georges {\it et al.} 1996)
to study a single electron 
in the Holstein model with binary disorder, 
focusing especially on polaron states at the 
high energy edge of the lowest polaronic 
subband. 
These states are extremely sensitive to disorder, with 
a return probability which, due to the phonon 
admixture, is several orders of magnitude larger
than for the states at the bottom of the subband. 
Within the DMFT it was however impossible to determine
the critical disorder needed to localize polaron states. 
Accordingly, mobility edges for polaron states could
not be determined.  

To overcome the limitations of the DMFT, we  
adopted therefore a recently proposed generalization of 
the DMFT, the statistical 
dynamical mean-field theory (statDMFT)~(Dobrosavljevi\'c and Kotliar 1997, 
Dobrosavljevi\'c and Kotliar 1998),
to the 
Holstein model with uniformly distributed on-site
energies~(Bronold and Fehske 2002, Bronold and 
Fehske 2003).
The statDMFT is not only capable of
accounting the polaron formation process but also the spatial 
fluctuations giving rise to Anderson localization. Within 
the statDMFT we could clearly distinguish between itinerant
and localized polaron states. The purpose of this paper is to 
give an extended account of our investigations, with a complete
description of the technical apparatus, including the details
of the numerical implementation of the statDMFT, and a full 
discussion of the localization properties of a single 
polaron.

As a generic model for a disordered polaronic material, we use the
Anderson-Holstein model (AHM) 
\begin{eqnarray}
H=\sum_{i\sigma} \epsilon_i n_{i\sigma} 
-\sum_{i,j,\sigma} J_{ij} c_{i\sigma}^\dagger c_{j\sigma} 
+\Omega \sum_i b_i^\dagger b_i
-\sqrt{E_p\Omega}\sum_{i\sigma}(b_i + b_i^\dagger)n_{i\sigma},
\label{model}
\end{eqnarray}
where $c_{i\sigma}^\dagger (b_i^\dagger)$ are electron (phonon) 
creation operators, $n_{i\sigma}=c_{i\sigma}^\dagger c_{i\sigma}$ is 
the electron density on site $i$, the electron transfer
integral $J_{ij}=J$ for ($i$, $j$) next neighbour sites and
zero otherwise, 
$\Omega$ is the bare phonon energy 
($\hbar=1$), and $E_p$ is the polaron shift. 
The on-site energies $\{\epsilon_i\}$ are assumed to be
independent, identically distributed random variables
with a uniform distribution 
$p(\epsilon_i)=(1/\gamma)\theta(\gamma/2-|\epsilon_i|)$.   

The `polaron properties' of the AHM are governed by two 
parameter ratios: the adiabaticity $\alpha=\Omega/J$, 
indicating whether the polaron is light $\alpha\ll 1$ or 
heavy $\alpha\gg 1$~(Rashba 1982, Wellein and Fehske 1998),
and a dimensionless
electron-phonon coupling constant, $\lambda=E_p/2J$ or 
$g^2=E_p/\Omega$.  
Polaron formation sets in if both $\lambda>1$ and $g^2>1$.
The internal structure of the polaronic quasi-particle 
depends on $\alpha$, in particular the  
momentum dependent phonon admixture. 
It is quite different 
in the adiabatic ($\alpha\ll 1$), non-adiabatic ($\alpha\sim 1$), 
and anti-adiabatic ($\alpha\gg 1$) regimes. Another important
effect occurs in the intermediate coupling regime, where 
electron-phonon coupling initiates long-range tunneling
processes~(Fehske {\it et al.} 1997a, Fehske {\it et al.} 1997b).
Evidently, how disorder affects polaron 
states strongly depends on the polaron parameters. As a result, 
the `localization properties' are expected to be quite different 
in the various polaronic regimes. 

The strength of disorder
is specified by the width $\gamma$ of the distribution
for the on-site energy $\epsilon_i$. The structure of the 
AHM suggests to  
distinguish two regimes: The weakly disordered
Holstein regime, where $\gamma$ is small on the scale of the bare bandwidth,
and the strongly disordered 
Anderson regime, with $\gamma$ large on the scale of the width of the 
polaronic subbands. 

The organization of the paper is as follows.
Section II gives a general description of the statDMFT. 
As a guidance, we first put in Subsection II.A our work
and the statDMFT into perspective. For
electrons interacting with phonons, the basic equations
are derived in Subsection II.B.  
The statDMFT reduces for 
lattices with large coordination number to the DMFT. This
is shown in the appendix. For practical calculations it is
convenient to use a Bethe lattice. Subsection II.C
specifies therefore the basic equations for the Bethe lattice.
Subsection II.D finally introduces the localization
criterion we used to identify localized states. Numerical
results in the single particle sector are presented in 
Section III. After specializing the basic equations in 
Subsection III.A to the
case of a single particle at zero temperature, we first  
give in Subsection III.B results with electron-phonon 
coupling turned off.
The pure Anderson case serves two purposes: First, it 
demonstrates, for a well understood case, that the method
and the localization criterion work; 
second, it is used as a reference point for the analysis of 
the polaron case presented in Subsection III.C.
Finally, we conclude in Section IV with a short outlook.

\section{Statistical dynamical mean field theory}     

\subsection{General considerations}

Before we go into technical details, we put our work 
and the statDMFT into perspective. Self-trapping in a disordered
environment is one example of physical 
problems where interaction and disorder effects interfere
with each other. The most prominent example is 
electron transport in the impurity band of doped 
semiconductors. Especially near the 
metal-insulator transition, Coulomb interaction and 
disorder effects cannot be separated, and it is suspected 
that it is precisely the
interplay of both, which leads to new physical 
phenomena, for instance, to the formation of local 
moments~(Finkel'stein 1983, Castellani {\it et al.} 1984)
or to a new emerging magnetic 
phase~(Kirkpatrick and Belitz 1990).

There is no method which could be mechanically used to 
investigate disordered interacting systems beyond 
perturbation theory. In particular, the strongly disordered 
regime, where Anderson localization takes place, is beyond the
applicability of most theoretical techniques. Nevertheless, there  
have been various attempts to generalize methods, which
have been successfully employed to investigate Anderson 
localization of noninteracting electrons, in such a way that they 
can be also utilized in situations where interactions are
important. Most efforts have been directed towards  
electrons interacting via short- and/or long-range
Coulomb potentials, with a few exceptions 
dealing with electrons coupled to 
phonons~(Anderson 1972, Girvin and Jonson 1980, 
Cohen {\it et al.} 1983, Mueller and Thomas 1983, 
Kopidakis {\it et al.} 1996, Bronold {\it et al.}
2001, Bronold and Fehske 2002, Bronold and Fehske 2003).

After Anderson's original work~(1958)
and 
Mott's suggestion~(1968a, 1968b, 1976, 1981)
of its 
connection to transport properties in amorphous semiconductors, 
three major techniques have been cultivated to study Anderson localization
quantitatively in further detail: 
(i) direct numerical simulations [see, e.g., the
review articles by Thouless (1974), Brezini and Zerki (1992),
Kramer and MacKinnon (1993)],
(ii) diagrammatic techniques~[Vollhardt and W\"olfle~(1980), for recent studies consult 
also the review articles by Kramer and MacKinnon (1993) and Vollhardt and W\"olfle 
(1992)],
and (iii) field-theoretical approaches~[Wegner~(1976), for recent 
studies consult Kramer and MacKinnon~(1993)].
Subsequently, all 
three techniques have been applied to interacting systems as well.
Direct numerical simulations are conceptually the simplest 
approach. However, they are restricted to small system sizes and 
usually to a truncated configuration type treatment of the 
Coulomb interaction. To this category belong, for instance,
all attempts to 
study the localization of two interacting electrons in the 
background of a 
frozen Fermi sea~(von Oppen {\it et al.} 1996).
In the case of electron phonon interaction, 
the size of the phonon Hilbert space is the limiting factor.
Diagrammatic methods work usually only for weak 
interactions~(Lee and Ramakrishnan 1985).
In principle, strong interactions can be also treated with diagrammatic
techniques, if they are based on renormalized 
expansions, but the proliferation of diagrams makes the practical 
implementation very often rather complicated if not impossible. 
Field-theoretical techniques, in contrast, for instance based on the 
construction of an effective field theory for the 
low-energy, long-wavelength 
excitations~(Belitz and Kirkpatrick 1994, 
Kamenev and Andreev 1999),
albeit very promising, operate in abstract spaces, with 
order parameter functions whose physical content is sometimes hard 
to grasp.

Anderson (1958), on the other hand, used 
a completely different approach. Instead of focusing on 
the calculation of averaged correlation functions, he
emphasized that, in general,
average values are not representative, 
especially in the strongly 
disordered
regime. All variables of the theory, in particular, 
observables, should be characterized by distributions. 
Specifically, he examined, in probabilistic terms,
the convergence properties of the renormalized perturbation 
series for the local hybridization function 
appearing in a locator expansion of the local Green function. 
Towards that 
end, he gave asymptotic estimates for the higher order terms of 
the series. The mathematical arguments are quite involved and  
have been further clarified later, most notably, by 
Thouless~(1970, 1974)
and Economou and Cohen~(1972).
A simplified 
probabilistic approach, based on the self-consistent 
solution of the second order renormalized perturbation
series, started with 
Abou-Chacra, Anderson and Thouless~(1973).
It proved to be a very powerful tool, applicable not only to 
substitutionally~(Kumar {\it et al.} 1975, Brezini 1982, 
Brezini and Olivier 1983, Miller and Derrida 1993)
but also to 
topologically disordered
materials~(Heinrichs 1977, Fleishman and Stein 1979, 
Elyutin 1979, Elyutin 1981, Logan and Wolynes 1985,
Logan and Wolynes 1986, Logan and Wolynes 1987).

The flexibility of the simplified probabilistic 
analysis of the renormalized perturbation series
has been recently utilized by 
Dobrosavljevi\'c and Kotliar~(1997, 1998)
to analyse 
various physical properties of disordered electrons
with strong local electron-electron 
correlations~(Miranda and Dobrosavljevi\'c 2001a,
Miranda and Dobrosavljevi\'c 2001b, Aguiar {\it et al.}
2003).
They successfully combined the
dynamical mean field approximation for the 
description of (local) correlations with 
Abou-Chacra, Anderson, and Thouless's probabilistic 
treatment of the spatial fluctuations due to randomness.
The statDMFT maps the original lattice 
problem onto an ensemble of impurity 
problems, which is then analysed in terms of probability. 
In contrast to the field-theoretical approaches, the 
method is very intuitive and conceptually simple. Moreover,
it captures local interaction processes nonperturbatively. 
It is therefore capable of addressing issues beyond standard 
diagrammatic perturbation theories.  
Numerically it is somewhat involved, but not restricted 
to small system sizes as direct numerical simulations.  
The main drawback of the method is its restriction to
local, short-range interactions. In principle, 
long-range interaction processes could be incorporated,
but only at
the level of static molecular fields (for instance,
in the case of Coulomb 
interaction a Hartree term would appear), 
which is probably not sufficient for a complete 
investigation of the interplay between
long-range interactions and disorder. For the AHM, however, 
where the (electron-phonon) coupling is local, the 
statDMFT's problem to deal with long-range interactions
does not occur and the statDMFT is expected to  
capture the essential physics of the interplay between
Anderson localization and self-trapping.

\subsection{Derivation of the basic equations}

The derivation of the statDMFT for a generic model 
of interacting electrons has been given  
by Dobrosavljevi\'c and Kotliar~(1997, 1998).
It is based on the cavity method and applies to arbitrary 
lattices, temperatures, and densities.
To make our presentation self-contained and to fix our 
notation, we outline in this subsection the cavity method,
carefully paying attention to the probabilistic interpretation
of the statDMFT and to specifics due to electron-phonon coupling. 

The starting point is the partition function 
\begin{eqnarray}
Z=\int\prod_{i,\sigma} {\cal D}c_{i\sigma}^\dagger{\cal D}c_{i\sigma} 
{\cal D}b_i^\dagger{\cal D}b_i e^{-S[c^\dagger,c,b^\dagger,b]},
\end{eqnarray}
where the action, for a fixed realization of the on-site energies, 
is given by 
\begin{eqnarray}
S[c,c^\dagger,b,b^\dagger]=\int_0^\beta d\tau 
\sum_{j,\sigma}\bigg\{c_{j\sigma}^\dagger(\tau)[\partial_\tau -\mu]
c_{j\sigma}(\tau)
+ b_j^\dagger(\tau)\partial_\tau b_j(\tau) 
+H[c,c^\dagger,b,b^\dagger](\tau)\bigg\}.
\end{eqnarray}

All sites but one are now integrated out, yielding   
an effective single-site action. For that purpose, the 
action is split into three parts: 
$S=S_i+S^{(i)}+\Delta S(i)$, where $S_i$ is the action of the
isolated site $i$, 
$S^{(i)}$ is the action for the system with site $i$ excluded, 
and 
\begin{eqnarray}
\Delta S(i)=-\int_0^\beta d\tau \sum_{l,\sigma}
\bigg\{c_{l\sigma}^\dagger(\tau)\eta_{l\sigma}(\tau) + 
\eta_{l\sigma}^\dagger(\tau)c_{l\sigma}(\tau)\bigg\},
\end{eqnarray}
with $\eta_{l\sigma}=J_{li}c_{i\sigma}$,
is the part of the action which connects site $i$ to its neighbours.

Identifying $\Delta S(i)$ as a source term to generate cavity
Green functions, i.e. Green functions for the lattice with 
site $i$ removed, the effective single-site action $S(i)$, 
defined by  
\begin{eqnarray}
e^{-S(i)}={1\over Z}
\int\prod_{j\neq i,\sigma} 
{\cal D}c_{j\sigma}^\dagger{\cal D}c_{j\sigma} {\cal D}b_j^\dagger
{\cal D}b_j e^{-[S_i+S^{(i)}+\Delta S(i)]},  
\end{eqnarray} 
can be written as 
\begin{eqnarray}
S(i)=S_0+S_i[c_{i\sigma}^\dagger,c_{i\sigma},b_i^\dagger,b_i]
-W^{(i)}[\eta^\dagger,\eta],  
\end{eqnarray}
where we explicitly indicated the fields on which the various
terms depend. The first term
$S_0$ is a constant. While the second term, the action for the
isolated site $i$, contains only the fields on site $i$, the third term,
\begin{eqnarray}
W^{(i)}[\eta^\dagger,\eta]=1+\sum_n
\int_0^\beta d\tau_1 ...\int_0^\beta d\tau_n 
\int_0^\beta d\tau_1'...\int_0^\beta d\tau_n' 
A^{[2n]}(\tau_1,...,\tau_n'),
\label{Wi}
\end{eqnarray}
with 
\begin{eqnarray}
A^{[2n]}(\tau_1,...,\tau_n')=
(-)^n\sum_{l_1,\sigma_1} ... \sum_{l_n',\sigma_n'}
\eta_{l_1\sigma_1}^\dagger(\tau_1) ... \eta_{l_n\sigma_n}^\dagger(\tau_n)
G_{l_1\sigma_1...l_n'\sigma_n'}^{(i)}(\tau_1, ... ,\tau_n')
\eta_{l_1'\sigma_1'}(\tau_1') ... \eta_{l_n'\sigma_n'}(\tau_n'),
\end{eqnarray}
involves fields from all sites.

The effective single-site action $S(i)$ contains the full information
of the original lattice action. Integrating out 
all sites but one introduces a
$2n-$point cavity Green function 
$G^{(i)}_{l_1\sigma_1...l_n'\sigma_n'}(\tau_1,...,\tau_n')$.
This function is extremely involved and approximations 
are required to obtain feasible 
equations. A crucial 
simplification can be made by truncating Eq. (\ref{Wi}) after   
the $n=1$ term. This leads to a quadratic effective single-site 
action, similar to the DMFT~(Georges {\it et al.} 1996), 
where it was
further shown that the truncation becomes exact for lattices with
infinite coordination number. As a consequence, although
the cavity construction is performed for lattices with 
a finite coordination number, interaction processes, here
due to the electron-phonon coupling, are treated as if the 
lattice had infinite coordination number.  As a result of 
the truncation, the statDMFT becomes a mean field theory.

Keeping only the $n=1$ term, Eq. (\ref{Wi}) depends only on
the electronic fields on site $i$ and reduces to  
\begin{eqnarray}
W^{(i)}[c^\dagger_{i\sigma},c_{i\sigma}]
=1-\int_0^\beta d\tau\int_0^\beta d\tau'\sum_\sigma 
c_{i\sigma}^\dagger(\tau)
H_{i\sigma i\sigma}(\tau-\tau')
c_{i\sigma}(\tau'), 
\end{eqnarray}
with the hybridization function
\begin{eqnarray}
H_{i\sigma i\sigma}(\tau-\tau')=J^2\sum_{lm}
G_{l\sigma m\sigma}^{(i)}(\tau-\tau'). 
\label{Hybrid}
\end{eqnarray}
The information of the 
lattice is now contained in the two-point cavity Green function 
$G_{l\sigma m\sigma}^{(i)}(\tau-\tau')$, describing all 
paths from site $l$ to site $m$ without passing through 
site $i$ (note that due to the definition of $J_{lm}$,
both sites $l$ and $m$ are neighbouring sites
to $i$). 

Absorbing irrelevant constants, which do not affect the
electron dynamics, into $Z_0$, the 
effective single-site action reads 
\begin{eqnarray}
S(i)&=&\int_0^\beta d\tau d\tau'\sum_\sigma
c_{i\sigma}^\dagger(\tau)
\bigg\{[\partial_\tau+\epsilon_i-\mu]\delta(\tau-\tau')
+H_{i\sigma i\sigma}(\tau-\tau')\bigg\}
c_{i\sigma}(\tau')
\nonumber\\
\nonumber\\
&+& S_{ph}(i) + S_{int}(i), 
\label{Seff}
\end{eqnarray}
with
\begin{eqnarray}
S_{ph}(i)=\int_0^\beta d\tau
b_i^\dagger(\tau)[\partial_\tau+\Omega]b_i(\tau) 
\end{eqnarray}
and 
\begin{eqnarray}
S_{int}(i)=-\int_0^\beta d\tau
\sqrt{E_p\Omega}[b_i(\tau)+b_i^\dagger(\tau)]n_{i\sigma}(\tau).    
\label{Sint}
\end{eqnarray}
According to Eq. (\ref{Seff}), 
the interaction between electrons and phonons takes place only on
site $i$, which, due to the hybridization function
$H_{i\sigma i\sigma}(\tau)$, is however
embedded into the full lattice. 

Although, for a given $H_{i\sigma i\sigma}(\tau)$, 
any local Green function could be obtained from $S(i)$, the set of 
equations is not closed. This can be most clearly seen from the
identity~(Dobrosavljevi\'c and Kotliar 1997,
Dobrosavljevi\'c and Kotliar 1998)
\begin{eqnarray}
\int_0^\beta d\tau' G_{l\sigma i\sigma}(\tau-\tau')
G_{i\sigma m\sigma}(\tau')=
\int_0^\beta d\tau' 
\bigg\{G_{l\sigma m\sigma}(\tau-\tau')-
G_{l\sigma m\sigma}^{(i)}(\tau-\tau')\bigg\}
G_{i\sigma i\sigma}(\tau'),
\label{identity}
\end{eqnarray}  
which connects the cavity Green function $G_{l\sigma m\sigma}^{(i)}(\tau)$
with the lattice Green function $G_{l\sigma m\sigma}(\tau)$.
Therefore, the hybridization function 
$H_{i\sigma i\sigma}(\tau)$ defined in Eq. (\ref{Hybrid}) is a 
functional of the full, nonlocal 
lattice Green function, which of course cannot be obtained from the 
effective single-site action $S(i)$. 

To close the set of equations, Dobrosavljevi\'c and 
Kotliar~(1997, 1998)
suggested to approximate in Eqs. (\ref{Hybrid}) and 
(\ref{identity}), for a given 
realization of disorder, the full lattice Green function 
$G_{l\sigma m\sigma}(\tau)$ by the bare lattice 
Green function $G^{0}_{l\sigma m\sigma}$,
with on-site energies shifted by the local self-energy.
In Matsubara space the  substitution reads 
\begin{eqnarray}
G_{i\sigma j\sigma}(i\omega_n)\rightarrow
G^{0}_{i\sigma j\sigma}
(i\omega_n)\bigg|_{\epsilon_i\rightarrow\epsilon_i+\Sigma_{i\sigma}(i\omega_n)},
\label{close}
\end{eqnarray}
with $\Sigma_{i\sigma}(i\omega)$ defined by  
\begin{eqnarray}
G_{i\sigma i\sigma}(i\omega_n)=
{1\over{i\omega_n-\epsilon_i+\mu-H_{i\sigma i\sigma}(i\omega_n)-
\Sigma_{i\sigma}(i\omega_n)}},
\label{Dyson}
\end{eqnarray}
which is the Fourier transformed Dyson equation 
for the local Green function 
$G_{i\sigma i\sigma}(\tau)=-
\langle T_\tau c_{i\sigma}(\tau)c_{i\sigma}^\dagger(0)\rangle_{S(i)}$
as obtained from the effective single-site action $S(i)$. 
Note, the substitution Eq. (\ref{close})
is not exact, because in general the interaction self-energy is  
nonlocal. Besides the  truncation of the effective single-site 
action, this is the second major approximation invoked by the
statDMFT.

Now the set of Eqs. (\ref{Hybrid})--(\ref{Dyson}) 
is closed and constitutes the basic statDMFT equations, 
applicable to arbitrary temperatures, densities, and lattices.
A self-consistent solution yields the local Green function 
$G_{i\sigma i\sigma}(i\omega_n)$. 

It is crucial to realize that, due to the randomness 
of the on-site energies $\epsilon_i$,  
the local Green function $G_{i\sigma i\sigma}(\tau)$,
the local self-energy
$\Sigma_{i\sigma}(\tau)$, and the local hybridization function
$H_{i\sigma i\sigma}(\tau)$ are random variables. More 
specifically, interpreting the site indices as labels
enumerating the elements of random samples, 
$G_{i\sigma i\sigma}(\tau)$, $\Sigma_{i\sigma}(\tau)$, and
$H_{i\sigma i\sigma}(\tau)$ can be understood as particular 
realizations of the random variables
$G_{\sigma}^{loc}(\tau)$,
$\Sigma_{\sigma}^{loc}(\tau)$, and
$H_{\sigma}^{loc}(\tau)$, respectively.
Thus, equations (\ref{Hybrid})--(\ref{Dyson}) 
are in fact stochastic recursion relations, from 
which random samples
for the random variables 
$G_{\sigma}^{loc}(\tau)$,  
$\Sigma_{\sigma}^{loc}(\tau)$, and 
$H_{\sigma}^{loc}(\tau)$ can be 
constructed.  
The iterative solution of Eqs. 
(\ref{Hybrid})--(\ref{Dyson})  
is therefore equivalent to the calculation of the distribution 
for the random variables $G_{\sigma}^{loc}(\tau)$,
$\Sigma_{\sigma}^{loc}(\tau)$, and
$H_{\sigma}^{loc}(\tau)$.  

Starting from
an initial sample $\{H_{i\sigma i\sigma}(\tau)\}$ 
for the local hybridization function
$H^{loc}_\sigma(\tau)$, a
sample $\{G_{i\sigma i\sigma}(\tau)\}$ for the local Green function 
$G^{loc}_\sigma(\tau)$ is obtained by solving the quantum-mechanical
many-body problem defined by 
$S(i)$ [cf. Eqs. (\ref{Seff})--(\ref{Sint})]. 
The Dyson equation (\ref{Dyson}) for the local Green function
yields then a sample $\{\Sigma_{i\sigma}(\tau)\}$
for the local self-energy $\Sigma^{loc}_\sigma(\tau)$, 
from which a new sample for the
hybridization function is calculated from Eqs. (\ref{close}),
(\ref{identity}),
and (\ref{Hybrid}). Going through this loop iteratively, 
the random samples are successively updated until a fix 
point is reached.

The iteration is numerically very involved for Bravais 
lattices (for instance, simple cubic, face-centered cubic,
or body-centered cubic lattices),  
in particular the construction of the
lattice Green function through Eq. (\ref{close}). 
In practice, the statDMFT is therefore 
formulated on a Bethe lattice, which is an infinite, loop-free
graph with connectivity $K$ (where $K+1$ is the number of 
next neighbours). Besides the reduced 
numerical complexity, the Bethe lattice has the 
additional advantage that the statistical dependence of
the random variables can be analysed in detail. 

Finally, we mention that the statDMFT reduces to the DMFT 
for lattices with 
large coordination number. For the particular case of a 
Bethe lattice, this is shown in the appendix. 

\subsection{Specification to a Bethe lattice}

\begin{figure}[t]
\hspace{0.0cm}\psfig{figure=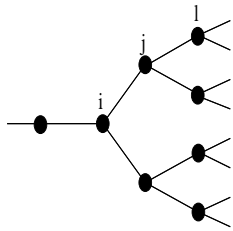,height=3.0cm,width=3.0cm,angle=0}
\caption[fig3]
{$K=2$ Bethe lattice.}
\label{fig3}
\end{figure}

Due to the absence of closed loops, the hybridization function 
on the Bethe lattice depends only on the {\it local} cavity Green 
function, which can be again obtained from an appropriate
effective single-site action. As a result, the statDMFT 
simplifies enormously on a Bethe lattice.  

Specifically for the 
$K=2$ Bethe lattice with nearest 
neighbour coupling shown in Fig. {\ref{fig3}, the hybridization 
function becomes 
\begin{eqnarray}
H_{i\sigma i\sigma}(i\omega_n)=J^2\sum_{j~{\rm NN}~i}
G_{j\sigma j\sigma}^{(i)}(i\omega_n),
\label{HybridBL}
\end{eqnarray}   
where now only local cavity Green functions appear. This is 
the crucial simplification because $G_{j\sigma j\sigma}^{(i)}(i\omega_n)$
can be obtained from the effective single-site action $S^{(i)}(j)$, 
which has the same structure as $S(j)$, except that the hybridization
function is now given by
\begin{eqnarray}
H_{j\sigma j\sigma}^{(i)}(i\omega_n)=J^2\sum_{l\neq i~{\rm NN}~j}
G_{l\sigma l\sigma}^{(ji)}(i\omega_n).
\label{forward}
\end{eqnarray}  
In addition, on a Bethe lattice, because of the absence of  
closed loops, the exclusion of site $j$ 
already ensures that site $i$ is not visited, i.e., 
$G_{l\sigma l\sigma}^{(ji)}(i\omega_n)=G_{l\sigma l\sigma}^{(j)}(i\omega_n)$.
Accordingly, except for the `base' site $i$, the hybridization function
for all other sides is simply given by the sum over the `forward' 
cavity Green function.
For the base site, in contrast, all the neighbouring cavity 
Green functions contribute [see Eq. (\ref{HybridBL})].  

As a consequence, taking $i$ as the base site, from which 
the cavity construction starts, gives for the 
Dyson equation for the local base site Green function 
\begin{eqnarray}
G_{i\sigma i\sigma}(i\omega_n)&=&
\left[
i\omega_n+\mu-\epsilon_i-J^2\sum_{j~{\rm NN}~i}
G_{j\sigma j\sigma}^{(i)}(i\omega_n)-\Sigma_{i\sigma}(i\omega_n)
\right]^{-1},
\label{Gii1}
\end{eqnarray}
whereas the Dyson equation for the local Green function at 
all other sites reads
\begin{eqnarray}
G_{j\sigma j\sigma}^{(i)}(i\omega_n)&=&
\left[
i\omega_n+\mu-\epsilon_j-J^2\sum_{l\neq i~{\rm NN}~j}
G_{l\sigma l\sigma}^{(j)}(i\omega_n)-\Sigma_{j\sigma}^{(i)}(i\omega_n)
\right]^{-1}.
\label{Gjji}
\end{eqnarray}      

Note, all cavity Green functions $G_{j\sigma j\sigma}^{(i)}(i\omega_n)$
obey Eq. (\ref{Gjji}); hence they are identically distributed.
Moreover, because the cavity self-energy 
$\Sigma_{j\sigma}^{(i)}(j\omega_n)$ is obtained from $S^{(i)}(j)$,
containing only `forward' sites with respect to site $j$, 
the cavity Green function are also statistically independent  
and $G_{j\sigma j\sigma}^{(i)}(i\omega_n)$ can 
be interpreted as a particular realization of the
random variable $G_\sigma^{cav}(i\omega_n)$.  
The statistical 
dependence between cavity Green functions at different energies
is another issue. Naturally, it depends on the interaction 
processes and the particular approximation employed to calculate 
the local self-energy (see below).

In principle the base site self-energy 
$\Sigma_{i\sigma}(i\omega_n)$ and the cavity  self-energy 
$\Sigma_{j\sigma}^{(i)}(i\omega_n)$ are different. 
If, in the spirit of the DMFT, we neglect however this difference and identify 
$\Sigma_{i\sigma}(i\omega_n)$ with $\Sigma_{j\sigma}^{(i)}(i\omega_n)$,  
Eq. (\ref{Gii1}) reduces for $i$ and $j$ next neighbour sites to
\begin{eqnarray}
G_{i\sigma i\sigma}(i\omega_n)=
{1\over{[G_{i\sigma i\sigma}^{(j)}(i\omega_n)]^{-1}
-J^2G_{j\sigma j\sigma}^{(i)}(i\omega_n)}}~. 
\label{Gii2}
\end{eqnarray}
Within this  approximation, the local Green function 
$G_{i\sigma i\sigma}(i\omega_n)$ is expressed in 
terms of two statistically
independent random variables,
$G_{i\sigma i\sigma}^{(j)}(i\omega_n)$
and $G_{j\sigma j\sigma}^{(i)}(i\omega_n)$. 
The identification of $\Sigma_{j\sigma}(i\omega_n)$ with
$\Sigma_{j\sigma}^{(i)}(i\omega_n)$ is on the Bethe lattice
analogous to the substitution Eq. (\ref{close}) for Bravais 
lattices. 
Even on the Bethe lattice, the  
statDMFT involves therefore two approximations: the truncation
of the effective single-site action and the conflation of the
interaction self-energies [reminiscent of the substitution 
Eq. (\ref{close}) for Bravais lattices].

\subsection{Localization criterion} 

A natural measure of the itinerancy suggested by the effective 
single-site
action $S(i)$ is the total tunneling rate from a given site $i$ defined by 
the imaginary part of the hybridization function,
$\Gamma_{i\sigma}(\omega)
=J^2\sum_{j~\mbox{\rm NN}~i}N^{(i)}_{j\sigma}(\omega)$,
where $N^{(i)}_{j\sigma}(\omega)$ is the local cavity density of states
(LCDOS), i.e., the imaginary part of the
cavity Green function $G_{j\sigma j\sigma}^{(i)}(\omega)$.
Obviously, a finite tunneling rate $\Gamma_{i\sigma}(\omega)$ 
implies an
extended state at energy $\omega$. Localized states, in contrast, 
lead to a vanishing tunneling rate. 
The tunneling rate vanishes if the LCDOS $N^{(i)}_{j\sigma}(\omega)$
vanishes, which, in this sense, is a transport quantity, and 
could by itself be used as a kind of `order parameter' for 
localization. 
The tunneling rate $\Gamma_{i\sigma}(\omega)$ as well as the
LCDOS $N_{j\sigma}^{(i)}(\omega)$ are random variables and whether
they vanish or not depends on their distribution. The
shape of the distributions has to change therefore dramatically
at the localization transition.

To understand why the shape of the distribution has to change, it
is convenient to look at the distribution of a related quantity,
the local density of states (LDOS), defined by 
\begin{eqnarray}
N_{i\sigma}(\omega)=-{1\over\pi}{\rm Im}G_{i\sigma i\sigma}(\omega),
\end{eqnarray} 
because it directly reflects the spatial dependence of the wave 
function: For an extended state at energy
$\omega$, where the weight of the wave functions is more or less
the same on every site, the distribution of the LDOS at energy
$\omega$ is symmetric and the most probable value   
coincides with the arithmetic mean value (average value).
Localized states with energy $\omega$, on the other hand, have
substantial weight only on a few sites. The distribution is
therefore extremely asymmetric, with a most probable value
much smaller than the arithmetic mean value, which is therefore not 
representative anymore. Since the LDOS $N_{j\sigma}(\omega)$
is closely related to the LCDOS $N_{j\sigma}^{(i)}(\omega)$,
the distributions for the LCDOS  
and the tunneling rate undergo the same 
characteristic change at the localization transition as the
distribution for the LDOS.  

A priori it is not clear by what moment (or moments) 
the asymmetric distributions should be characterized. 
At this point it is useful to anticipate that
distributions, in particular the distribution for the local 
hybridization function,
can be used to calculate averaged
transport quantities, such as the dc conductivity or the return
probability (see below). 
The most probable value, i.e., the maximum of the
distribution, is then expected to play a crucial role. [Note, 
averaged four-point functions (transport quantities) 
capture the localization effect, 
in contrast to averaged two-point functions.]
It would be therefore natural to focus on the most probable 
values. The drawback is, however, that they 
cannot be directly obtained from the random samples; they 
require the explicit construction of histograms. 
More convenient quantities seem to be the so-called typical 
values~(Dobrosavljevi\'c and Kotliar 1997,
Dobrosavljevi\'c and Kotliar 1998),
which are simply the geometric
mean values of the random samples and yet
capture the asymmetry of the distributions reasonably 
well.  
Therefore, we focus in this paper on the typical values.         

In particular, we consider the distribution of 
the LDOS and distinguish localized from extended states 
by a detailed investigation of the average LDOS, 
\begin{eqnarray}
N_{\sigma}^{\rm ave}(\omega)=
{1\over N_s}\sum_i N_{i\sigma}(\omega),
\end{eqnarray}  
and the typical LDOS 
\begin{eqnarray}
N_{\sigma}^{\rm typ}(\omega)=
\exp\bigg[{{1\over N_s}\sum_i \log{N_{i\sigma}(\omega)}}\bigg].
\end{eqnarray}   
Specifically, we classify states at energy $\omega$ with 
$N_{\sigma}^{\rm ave}(\omega)\neq 0$ 
as localized if $N_{\sigma}^{\rm typ}(\omega)\rightarrow 0$ and
as extended if $N_{\sigma}^{\rm typ}(\omega)\neq 0$. Note that the LDOS 
is defined for real energies. The numerical solution has to 
take this into account (see below). It should be stressed,
Mirlin and Fyodorov~(1994)
also used the LDOS as an order parameter for the Anderson transition. 

\section{Numerical results}

\subsection{StatDMFT equations for a single electron} 

This section presents a detailed numerical investigation of a single
electron in the AHM at zero temperature. We solve the 
statDMFT equations on a Bethe lattice 
where, as mentioned in the previous
section, the numerical effort is manageable.  
As in other areas of statistical physics, the 
Bethe lattice calculation has 
the status of a mean field calculation for Bravais lattices.
We expect therefore our results to be qualitatively 
valid for any Bravais lattice with $d\ge 3$.

The easiest way to obtain the zero temperature, single-particle limit 
of the  statDMFT equations 
[Eqs. (\ref{Hybrid})--(\ref{Dyson})] is to work with 
an effective single-site 
Hamiltonian, instead of an effective single-site action. Neglecting
the spin, which for a single electron is irrelevant, 
the effective Hamiltonian reads
\begin{eqnarray}
H(j)^{(i)}=\epsilon_j c_j^\dagger c_j+\Omega b_j^\dagger b_j
-\sqrt{E_p\Omega}(b_j^\dagger+b_j)c_j^\dagger c_j
+\sum_\nu E_\nu a_\nu^\dagger a_\nu
+\sum_{\nu}[T_{\nu j}a_\nu^\dagger c_j + H.c.],
\label{Heff}
\end{eqnarray} 
where the notation indicates that the 
Hamiltonian models the dynamics encoded in the cavity 
effective 
single-site action $S(j)^{(i)}$. 

The Hamiltonian representation of $S(j)^{(i)}$
is not unique. In Eq. (\ref{Heff}) the embedding is given 
by the hybridization with an auxiliary field described by the
operator $a_\nu$. The 
auxiliary parameters $E_{\nu}$ and $T_{\nu j}$ have no physical 
meaning, they just parameterize the hybridization function
in terms of a spectral representation:  
\begin{eqnarray}
H_{jj}(i\omega_n)=
\sum_\nu{{|T_\nu|^2}\over{i\omega_n-\mu-\epsilon_j-E_\nu}}~.
\end{eqnarray}

For a given set of parameters $\{E_\nu, T_{\nu j}\}$, 
i.e. hybridization function $H_{jj}(i\omega_n)$, the 
local self-energy due to electron-phonon coupling 
$\Sigma_{i}(i\omega_n)$ can be deduced from
Eq. (\ref{Heff}). For finite densities, this can be 
done only by some approximation, introducing 
further uncertainties into the approach.   
For a single electron, however, the interaction 
self-energy can be obtained exactly 
in the form of a continued 
fraction~(Sumi 1974, Ciuchi {\it et al.} 1997),
valid over the whole range of parameters. Thus, 
for a single electron, no further approximations  
are required.

To be more specific, we calculate from Eq.(\ref{Heff})
the single-electron Green function for $T=0$, 
using the continued fraction 
expansion for the electron-phonon self-energy, 
identify then this Green function with
the zero temperature, single-electron limit of the cavity 
Green function in Eq. (\ref{Gjji}). 
Setting $i\omega_n+\mu\rightarrow z=\omega+i\eta$, we obtain  
\begin{eqnarray}
G^{(j)}_{ii}(z)=
\frac{1}{\displaystyle [F_{ii}(z)]^{-1}-
\frac{\Omega E_p}{\displaystyle [F_{ii}(z-1\Omega)]^{-1}-
\frac{2\Omega E_p}{\displaystyle [F_{ii}(z-2\Omega)]^{-1}-
\frac{3\Omega E_p}{...}}}},  
\label{Giij}
\end{eqnarray}
with  
\begin{eqnarray}
F_{ii}(z)=
\left[z-\epsilon_i-J^2\sum_{j}G_{jj}^{(i)}(z)\right]^{-1}.
\label{Fii}
\end{eqnarray}
The local Green function follows from Eq. (\ref{Gii2}),
setting again $i\omega_n+\mu\rightarrow z=\omega+i\eta$, 
\begin{eqnarray}
G_{ii}(z)=
{1\over{[G_{ii}^{(j)}(z)]^{-1}
-J^2G_{jj}^{(i)}(z)}}~.
\label{Gii3}
\end{eqnarray}

According to the interpretation given in the previous section, 
Equations (\ref{Giij})--(\ref{Gii3}) constitute stochastic 
recursion relations for a random sample 
for the local Green function $G_\sigma^{loc}(z)$. 
Note, Green functions with shifted energy appear in the 
continued fraction. They satisfy the same recursion relations,
with energy shifted to $z-q\Omega$. 
 
For a fixed energy $z-q\Omega$, the local Green functions 
$G_{ii}(z-q\Omega)$ are identically distributed.  
They are however not independent. As a consequence, 
for each energy $z-q\Omega$, we have to construct a 
separate random sample 
$\{G_{jj}^{(i)}(z-q\Omega)\}$ (with $j=1,...,N_s$ and $q=0,1,...,M$, where
$N_s$ and $M$ are the sample size and the maximum depth of the 
continued fraction, respectively), 
starting 
from an initial sample, which we successively update via a Monte Carlo 
algorithm, similar to the schemes described by other 
authors~(Abou-Chacra {\it et al.} 1973, Girvin and Jonson 1980),
drawing the random variables on the rhs of Eq. (\ref{Giij})
from the corresponding random samples created by the iteration step before. 
Iterating this process a sufficient number of times yields, as a fix 
point of the 
stochastic recursion relations, a self-consistent random sample for the 
cavity Green function at energy $z-q\Omega$. The  self-consistent 
random samples are then used to directly determine average and typical 
values of the respective random variables, or, in the form of 
histograms, the distributions associated with them.  

For reasonable numerical accuracy, the sample size $N_s$, which should not be
confused with the actual size of the Bethe lattice, but instead  
gives the precision with which we construct the random sample 
(cf. distributions), has to be sufficiently
large. Typical sample sizes are $N_s\approx 50~000$. The
maximum depth $M$ of the continued fraction, which describes the
maximum number of virtual phonons in the lattice, has to be large enough in 
order to capture polaron formation. As a rough estimate we use 
$M\approx 5g^2$, where 
$g^2=E_p/\Omega$ is approximately the average number of virtual 
phonons comprising the 
phonon cloud of the polaron.  

To distinguish localized from extended states, 
it is crucial to investigate the stochastic recursion relations in the limit
$\eta\rightarrow 0$~(Abou-Chacra {\it et al.} 1973).
Numerically this seems delicate. 
However, initializing the iteration loop with a finite imaginary 
part of the cavity Green functions, the stochastic recursion relations
can be iterated without problems for 
$\eta=0$~(Alvermann 2003).
The correct,
self-consistent value of the imaginary part of the cavity 
Green function is then realized during the iteration.  

Tracking random samples as a function of the iteration index
provides therefore information about the character of the states
at the energy for which the sample is constructed.
Random samples corresponding to extended states
converge rapidly to its fix point, whereas random samples
corresponding
to localized states do not converge. Hence, extended and
localized states can be detected by the differences they 
give rise to in
the mathematical properties of the recursion relations.

Sometimes it is advantageous
to keep $\eta$ finite. But then it is necessary to monitor the 
flow of the random samples, or of its associated 
average and typical values, as $\eta$ gets smaller and smaller.  
The part of the spectrum corresponding to extended states is 
insensitive to the
$\eta$-scaling, after $\eta$ is small enough to see the 
correct LDOS 
nothing changes anymore. Localized states on the
other hand are strongly affected. In fact, it can be 
numerically verified, that the typical LDOS indeed scales to
zero with 
$\eta\rightarrow 0$~(Miller and Derrida 1993, Alvermann 2003).

To investigate the localization properties of electron and polaron 
states, we specifically focused on the typical LDOS, tracking 
this quantity, for given energy and model parameters, 
as a function of $\eta$ and the number of iterations. 
Introducing a rescaled transfer amplitude 
$J=\bar{J}/\sqrt{K}$, we measure energies 
in units of the bare bandwidth $W_0=4\bar{J}=1$ and define 
$\bar{\lambda}=E_p/2\bar{J}$,  
$\bar{\alpha}=\Omega/\bar{J}$, and $\bar{\gamma}=\gamma/4\bar{J}$.      
All calculations, except in the appendix, are carried out for the
$K=2$ Bethe lattice.  

\subsection{Electron states}

\begin{figure}[t]
\hspace{0.5cm}\psfig{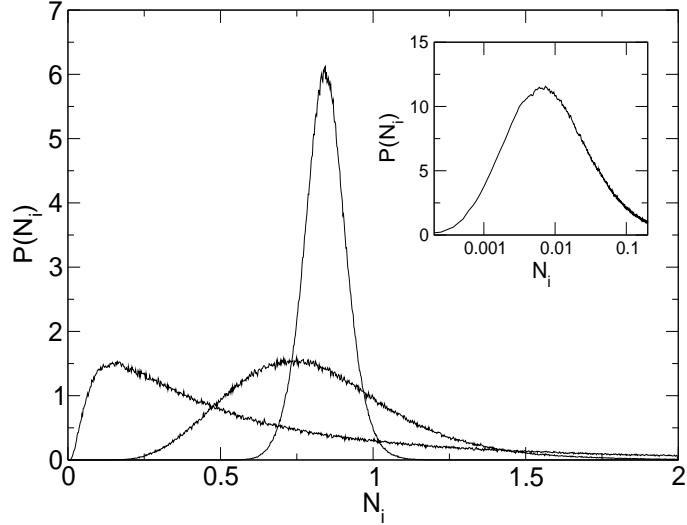}
\caption[fig5]
{Distribution for the LDOS at $\omega=0$ for
$\bar{\gamma}=0.2, 0.5$, and $1.0$.
The inset shows the
distribution for $\bar{\gamma}=1.5$ on a logarithmic scale.
In all cases $\eta=10^{-10}$.
}
\label{fig5}
\end{figure}
\begin{table}[t]
\caption[table1]
{Average, most probable, and typical LDOS
at $\omega=0$ corresponding to the distributions
shown in figure~\ref{fig5}.}
\begin{tabular}{cccc}
$\bar{\gamma}$ & $N^{ave}$ & $N^{mpv}$ & $N^{typ}$ \\ \hline
~0.2~ & ~0.844~ & ~0.843~ & ~0.841~ \\
~0.5~ & ~0.822~ & ~0.749~ & ~0.773~ \\
~1.0~ & ~0.710~ & ~0.162~ & ~0.445~ \\
~1.5~ & ~0.568~ & ~0.006~ & ~0.106~
\label{table}
\end{tabular}
\end{table}

To demonstrate the feasibility of the statDMFT and the associated  
localization criterion, we first discuss the localization
properties of a single electron without coupling to phonons.
In the next subsection, we will then contrast this well understood 
situation with the results for the coupled electron-phonon system.  
Some of the results of this subsection are taken from  
Alvermann~(2003).

We begin with a qualitative discussion of the distribution
for the LDOS at $\omega=0$, shown in figure \ref{fig5}
for $\bar{\gamma}=0.2, 0.5, 1.0$, and $1.5$ (inset). The shape of the
distribution changes dramatically with increasing 
disorder: For small disorder ($\bar{\gamma}<1$), the distribution is 
approximately
a Gaussian. The most probable value and the average value 
almost coincide and change little with disorder 
(see table~\ref{table}). Thus, the main effect of small 
disorder is to broaden the distribution.
For larger disorder ($\bar{\gamma}\ge 1$), 
on the other hand, the distribution develops a long tail, 
where the most probable  
value is now significantly smaller than the average 
value (see again table~\ref{table}). For even larger 
disorder, close to the localization transition 
(for instance, for $\bar{\gamma}=1.5$ 
shown in the inset of figure~\ref{fig5}), the distribution 
becomes almost log-normal~(Montroll and Shlesinger 1983).
The average LDOS is now
completely meaningless. Obviously, disorder does not
only broaden the distribution but induces, if  
sufficiently strong, a change in the nature of the
distribution.  
\begin{figure}[h]
\hspace{0.5cm}\psfig{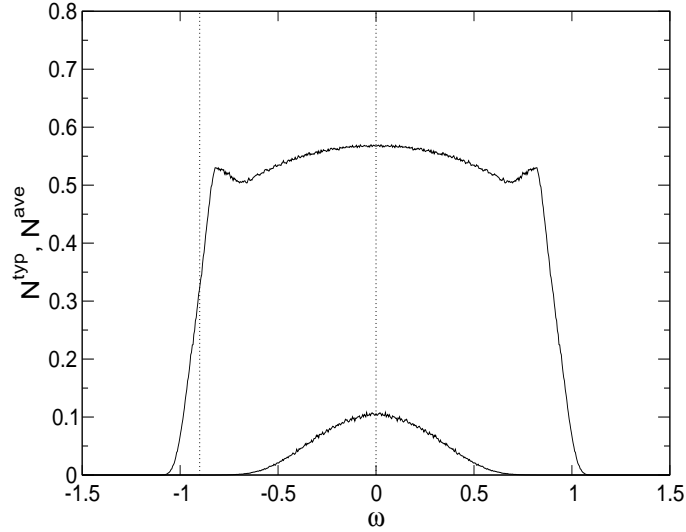}
\caption[fig6]
{Average (upper curve) and typical LDOS
(lower curve) for $\bar{\gamma}=1.5$ and $\eta=10^{-10}$. The
vertical dotted lines indicate two energies, one below and one
above the lower mobility edge.
}
\label{fig6}
\end{figure}
From physical considerations, it is clear,  
that the distribution of the LDOS for a fixed energy depends 
strongly on whether the states at that energy are 
extended or localized. 
The amplitude of 
an extended state is uniformly distributed over the 
whole system. Thus, the LDOS does not fluctuate much
from site to site, i.e., the distribution of the LDOS
at a given energy has to be centered around an 
average LDOS. 
On the other hand,
the amplitude of a localized state is essentially zero
everywhere, except in a small range around a central
site. For a given energy there is a certain number of
degenerate localized states, each centered at a different
site. The LDOS fluctuates therefore strongly throughout
the lattice. 
Accordingly, the distribution of the 
LDOS, for a given energy, has to be very broad, 
with a long tail, 
due to the few sites where localized states have  
a large weight. 
Naturally, a complete characterization of the LDOS
has to take the changes in the distribution into account,
especially near the localization transition.

Within the statDMFT, distributions are represented 
by random samples. 
Any moment of the distribution, as well as the 
most probable value can be obtained from the 
histogram associated with the random sample. 
As indicated in Section II.D, in practice it is often 
better to avoid the construction of the
histogram and to characterize, instead, physical 
quantities by their typical values, which are 
simply the geometric averages of the random 
samples. From Table \ref{table} we see that the 
typical values capture the characteristic 
asymmetry of the 
distribution in the localized regime reasonably 
well ($N^{typ}~\ll~N^{ave}$ for $\bar{\gamma}~>~1$). 
It is therefore indeed meaningful to use the typical 
LDOS as a kind of order parameter. 

Let us now turn 
to a quantitative analysis of our results.
In figure~\ref{fig6} we depict, for 
$\bar{\gamma}=1.5$, the typical and average LDOS over the 
whole spectral range of the pure Anderson model.
The two main effects of disorder, the appearance of Lifshitz
tails below  $\omega=\pm 1.0$ and the existence of 
mobility edges around $\omega \sim\pm 0.8$, characterized 
by a vanishing of the  typical LDOS (recall
subsection II.D), can be clearly
seen. The data have been obtained for $\eta=10^{-10}$, 
which is sufficiently small to reveal the intrinsic 
spectrum and to allow for a first estimate of 
the mobility edge. 

A precise quantitative 
determination of the position of the mobility edge, 
requires a calculation of the typical LDOS for 
$\eta\rightarrow 0$. We specifically
consider the lower mobility edge. With the 
obvious modifications all statements also hold 
for the upper mobility edge. 
\begin{figure}[h]
\hspace{0.5cm}\psfig{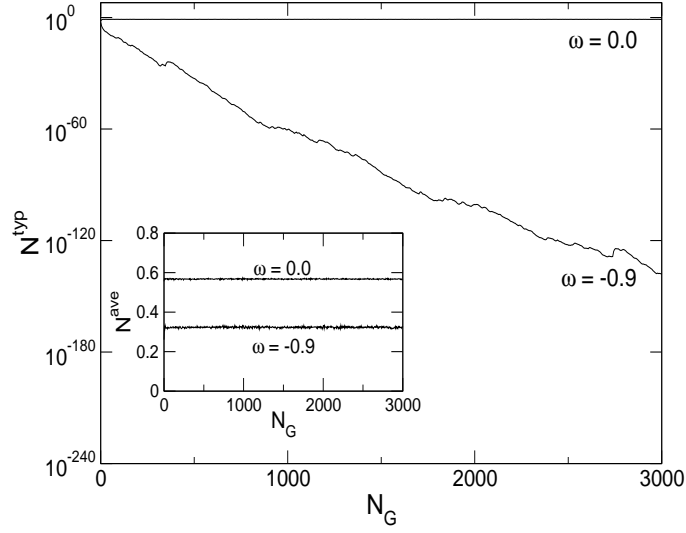}
\caption[fig7]
{Typical and average LDOS below ($\omega=-0.9$) and above ($\omega=0$)
the lower mobility edge for $\bar{\gamma}=1.5$ and $\eta=0$ as
a function of sample generation.
}
\label{fig7}
\end{figure}
To perform the $\eta\rightarrow 0$ limit 
numerically, we initialize
the sample with a finite LDOS and then iterate the
stochastic recursion relations for $\eta=0$ until 
convergence.
The iteration process yields the correct,
 self-consistent LDOS. This procedure works 
extremely well, as can be seen in 
figure~{\ref{fig7}, where we plot the typical and
average LDOS as a function of the iteration
step (in other words, the sample generation) 
for $\bar{\gamma}=1.5$,
$\eta=0$, and two energies, one below the
lower mobility edge ($\omega=-0.9$) and one above 
($\omega=0$). As expected, 
below the lower mobility edge, the self-consistent 
value of the typical LDOS 
continues to decrease with sample 
generation, whereas above the mobility edge, the
typical LDOS remains finite. Note, in both cases, the
average LDOS stays finite and is essentially independent
of the sample generation. This is the key observation, below
(above) the lower mobility edge, the typical LDOS vanishes (stays 
finite), whereas the average LDOS remains finite in both
cases.

At this point we should mention that the stable calculation of the 
average LDOS is subtle, because of the pure
statistics associated with the long tail, which, on the
other hand, determines the average value. To overcome
this problem, we calculated the average LDOS after each 
iteration step and then performed an arithmetic average 
over the obtained values of the average LDOS. Since 
the typical number of iterations $N_G\approx 1000$, we 
thereby effectively enlarged the sample size by three orders 
of magnitude, which was sufficient to obtain smooth 
data for the average LDOS~(Alvermann 2003).

Thus, the tracking of the typical LDOS as a function of 
the sample 
generation can be utilized to decide whether states for a 
given energy are localized or not. In practice one would 
screen all energies comprising the spectrum and set
a small threshold for the typical LDOS, for instance 
$10^{-30}$, below which the typical LDOS is assumed to 
be de facto zero. According to our localization 
criterion, states at that energy are then classified as 
localized. Since this procedure can be performed 
arbitrarily close to the mobility edge, the position of the 
mobility edge can be determined very precisely. It should 
be also mentioned that values for the typical LDOS of the order 
of $10^{-200}$ are not numerical artifacts. The small values are
correctly captured by the floating point representation of 
the typical LDOS.  

An alternative is to use finite 
values of $\eta$ and to track the typical LDOS with decreasing 
$\eta$. Below the lower mobility edge, the typical LDOS converges after
a sufficient number of iterations to a value whose scale 
is set by the chosen $\eta$ value,
whereas above the lower mobility edge, the typical  LDOS
reaches an intrinsic value, completely unrelated to $\eta$. 
The average LDOS is in both cases again finite.
For this procedure to work, $\eta$ has to be of course small
enough to resolve the intrinsic spectrum. For instance,
$\eta$ should not mimic Lifshitz tails.  
\begin{figure}[t]
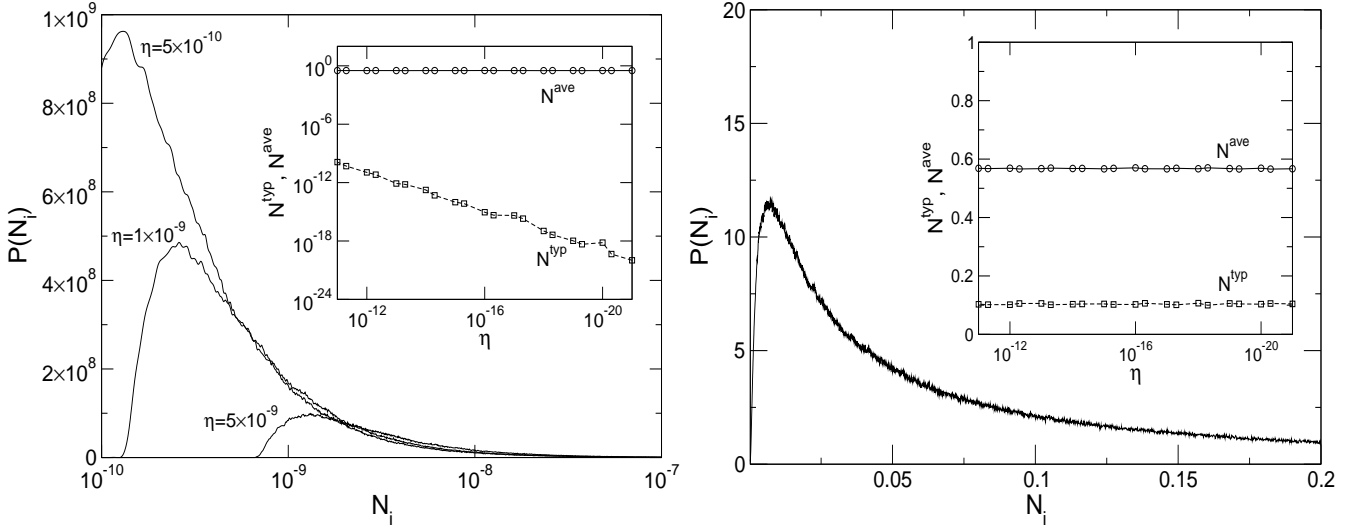

\begin{minipage}{0.5\linewidth}
\psfig{figure=Fig5a.eps,height=7.0cm,width=1.0\linewidth,angle=0}
\end{minipage}
\begin{minipage}{0.48\linewidth}
\psfig{figure=Fig5b.eps,height=7.0cm,width=1.0\linewidth,angle=0}
\end{minipage}
\caption[fig8]
{The left and right panels show, respectively, the $\eta$-scaling
of the distribution for the LDOS below ($\omega=-0.9$) and above
($\omega=0$) the lower mobility edge for $\bar{\gamma}=1.5$.
In the insets we display the
corresponding average and typical LDOS as a function of $\eta$.
}
\label{fig8}
\end{figure}
That the procedure based on the $\eta$-dependence works excellently, 
can be verified in figure~\ref{fig8}. To clarify the
origin of the 
differences in the $\eta$-scaling of the typical and average 
LDOS below and above the lower mobility edge, we first look at
the distributions. Below the lower
mobility edge at $\omega=-0.9$, 
shown in left panel of figure~\ref{fig8}, the distribution
changes radically with decreasing $\eta$ (note the 
logarithmic scale). The maximum of the distribution, 
i.e. the most probable value, shifts to very small values. 
In fact, $\eta$ sets the scale for the most probable value. 
Accordingly, the typical value shown in the inset, 
decreases with  
$\eta$. Note, however, the 
average value stays at a constant value, independent of 
$\eta$. Clearly, the average value is 
determined by the long tail, i.e., the few sites, 
where the localized wave functions have appreciable 
weight. The average LDOS is therefore extremely
less probable and indeed not a  
representative value. In contrast, above the lower mobility edge
at $\omega=0$,
shown in the right panel of figure~\ref{fig8}, the distribution is 
essentially independent of $\eta$. Accordingly, the typical
as well as the average LDOS do not change with
$\eta$. 

Below the mobility edge, the distributions for the LDOS, 
the LCDOS, and the tunneling rate $\Gamma^{loc}(\omega)$
become singular for $\eta\rightarrow 0$.
This behaviour is closely connected with 
the configuration averaged, spectrally resolved return probability 
\begin{eqnarray}
f_{ii}(\omega)=\lim_{\eta\rightarrow 0}
(\eta/\pi)\langle|G_{ii}(\omega+i\eta)|^2\rangle_{\{\epsilon_i\}},
\end{eqnarray} 
which is finite for localized states and zero for extended states. 
To reveal this relationship, we 
follow Logan and Wolynes~(1987)
and
introduce a joint distribution for 
the real and imaginary 
parts of the local hybridization function, 
$P(R(\omega),\Gamma(\omega))$, with 
$R(\omega)={\rm Re}H^{loc}(\omega)$ and 
$\Gamma(\omega)={\rm Im}H^{loc}(\omega)$, in terms of which we obtain
\begin{eqnarray}
f_{ii}(\omega)=\lim_{\eta\rightarrow 0}
\int_0^\infty d\Gamma(\omega)  
{{D(\omega,\Gamma(\omega))}\over{1+\eta^{-1}\Gamma(\omega)}}~,
\label{fii}
\end{eqnarray}
with
\begin{eqnarray}
D(\omega,\Gamma(\omega))={1\over\pi}\int_{-\infty}^\infty d\epsilon
\int_{-\infty}^\infty dR(\omega)
{{(\eta+\Gamma(\omega))p(\epsilon)
P(R(\omega),\Gamma(\omega))}
\over{[\omega-\epsilon-R(\omega)]^2+[\eta+\Gamma(\omega)]^2}}~.
\end{eqnarray}
If the state at energy $\omega$ is localized, the (marginal) distribution  
for $\Gamma(\omega)$, i.e. 
$P(\Gamma(\omega))=\int_{-\infty}^\infty dR(\omega)
P(R(\omega),\Gamma(\omega))$, is 
strongly peaked around 
$\Gamma^{mpv}(\omega)\rightarrow 0$ (cf. figure~\ref{fig8}). In that regime,  
$D(\omega,\Gamma(\omega))\approx
\delta(\Gamma(\omega)-\Gamma^{mpv}(\omega)) N^{cav}(\omega)$, with 
$N^{cav}(\omega)=-(1/\pi){\rm Im}G_{jj}^{(i)}(\omega)$ 
denoting the LCDOS~(Logan and Wolynes 1987).
Accordingly, Eq. (\ref{fii}) reduces to  
\begin{eqnarray}
f_{ii}(\omega)=\lim_{\eta\rightarrow 0}
{{N^{cav}(\omega)}\over{1+\eta^{-1}\Gamma^{mpv}(\omega)}},
\end{eqnarray}  
which, provided $N^{cav}(\omega)\neq 0$, is finite, 
because $1+\eta^{-1}\Gamma^{mpv}(\omega)\rightarrow 2$ 
for $\eta\rightarrow 0$. In other words, the singular behaviour of 
the distribution for $\Gamma^{loc}(\omega)$ gives rise to a pole in
$\langle|G_{ii}(\omega+i\eta)|^2\rangle_{\{\epsilon_i\}}$
which in turn makes $f_{ii}(\omega)$ finite. From diagrammatic 
approaches it is known that maximally crossed diagrams, 
responsible for destructive interference, produce a pole in 
$\langle|G_{ii}(\omega+i\eta)|^2\rangle_{\{\epsilon_i\}}$.  
Equation (\ref{fii}) provides therefore a link between 
the interference-based description of Anderson localization and
the probabilistic approach adopted by the statDMFT. 
\begin{figure}[t]
\hspace{0.5cm}\psfig{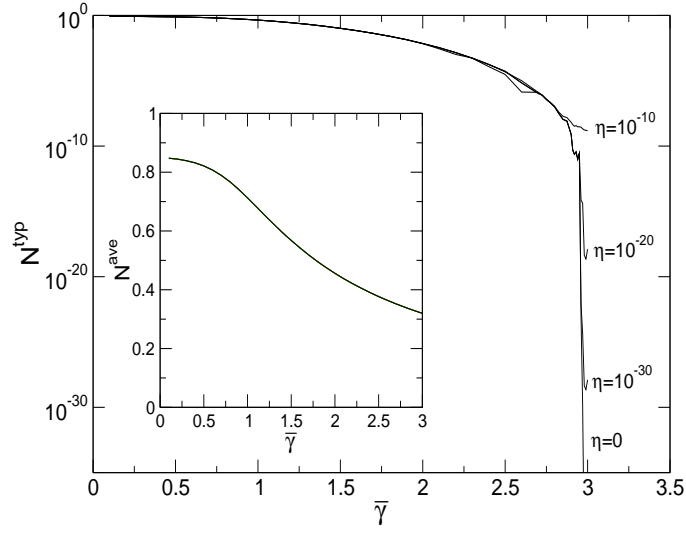}
\caption[fig9]
{$\eta$-scaling of the typical and average (inset)
LDOS at $\omega=0$ for $\bar{\gamma}=1.5$. Note, for the chosen
$\eta$ values, the average LDOS is
independent of $\eta$.
}
\label{fig9}
\end{figure}
The $\eta$-scaling discussed in figure~\ref{fig8} can be
followed all the way down to $\eta=0$, suggesting a 
precise procedure for the determination of mobility edges. 
In practice, however, we cut off the scaling at a certain 
threshold,
for instance $10^{-30}$. If the typical LDOS for a given
energy $\omega$
is below the threshold, and if for that energy the  
average LDOS is finite, we classify the state  
at energy $\omega$ as localized.     
For $\omega=0$ this is shown in figure~\ref{fig9}.
Clearly, for  $\omega=0$ the typical LDOS tends to   
zero for $\bar{\gamma}\sim 2.9$. The average LDOS on the other
hand is independent of $\eta$ and remains finite. 
Thus, according to our localization
criterion, the state at energy $\omega=0$ is localized
for $\bar{\gamma}\ge 2.9$, in accordance with
results obtained by Girvin and Jonson~(1980)
and by Miller and Derrida~(1993).

Performing the analysis for all energies comprising the spectrum, 
we can systematically map out the mobility edge trajectory. 
The result is shown in figure~\ref{fig10}.
Note that the statDMFT 
is accurate enough to detect the two transitions associated with 
the characteristic reentrance behaviour of the 
mobility edge trajectory near the band edge: The 
delocalization transition at small disorder and the 
localization transition at large disorder.   
Even in the strongly 
disordered regime, where the mobility edges move to the 
centre of the band and all states are localized, the statDMFT 
works reliably well. We did not attempt to estimate the
statistical error, but it should be of the order of 
the fluctuations visible in the plot.
\begin{figure}[t]
\hspace{0.5cm}\psfig{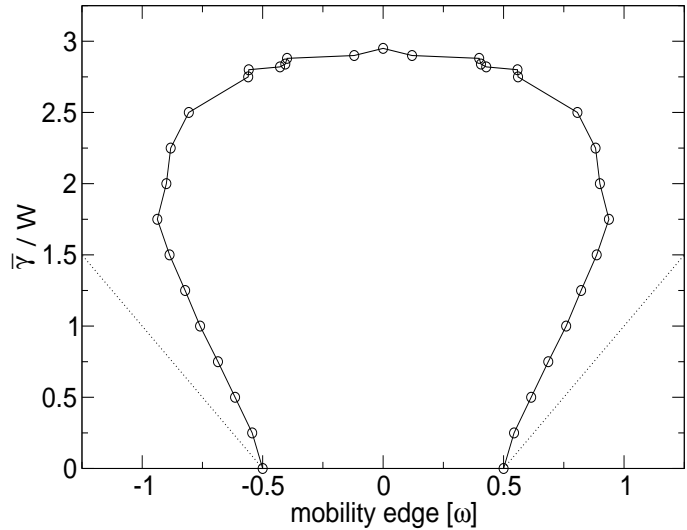}
\caption[fig10]
{Mobility edge trajectory for an electron in the pure
Anderson model. The dotted lines indicate the position of the band edges
defined by $\pm(W/2 +\gamma/2)$.
}
\label{fig10}
\end{figure}
Our results are in excellent agreement
with results obtained by Miller and 
Derrida~(1993),
suggesting that within the statDMFT mobility edges for 
noninteracting electrons can be indeed determined. In
the next subsection, we shall demonstrate, that the 
statDMFT is flexible and powerful 
enough, to determine mobility edges in interacting
systems as well.  

\subsection{Polaron states}

We now turn our attention to the consequences of the electron-phonon 
coupling. If the electron-phonon coupling is weak, the main effect 
is that the electron states above the phonon emission threshold,
i.e., states whose kinetic energy is at least the phonon
energy $\Omega$, acquire a finite lifetime due to 
inelastic scattering, which gives rise to ${\rm Im}\Sigma_i(\omega)\neq 0$. 
As a result, the critical disorder needed to localize these 
states will be larger than without electron-phonon 
coupling.
For the states below the phonon emission threshold, we have
${\rm Im}\Sigma_i(\omega)=0$, and the critical disorder is the same 
as for the  respective states in the pure 
Anderson model~(Bronold and Fehske 2002).  
\begin{figure}[h]
\hspace{0.5cm}\psfig{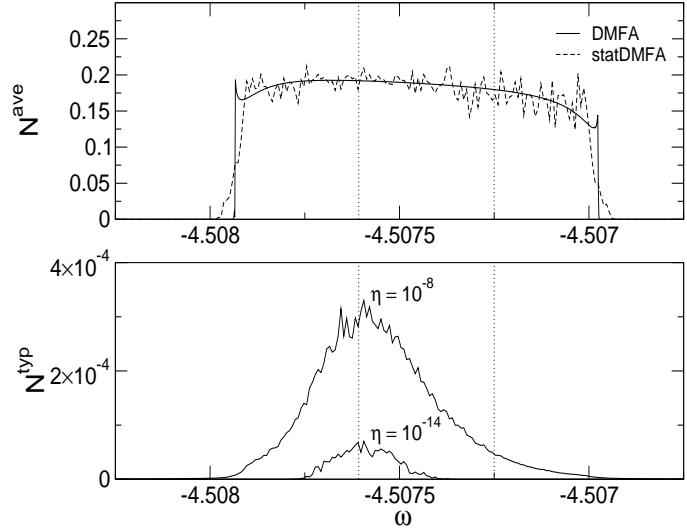}
\caption[fig11]
{
Average and typical LDOS for the lowest polaronic subband in the
anti-adiabatic, strong coupling regime for $\bar{\gamma}=2.5\times W$;
$W=3.45\times 10^{-4}$ is the width of the lowest  subband of the
pure Holstein model in units of
$W_0=4\bar{J}=1$.
The polaron parameters are $\bar{\alpha}=2.25$ and $\bar{\lambda}=9$.
The vertical dotted lines
indicate two energies, one above ($\omega=-4.50725$)
and one below ($\omega=-4.50759$) the upper mobility
edge for the subband.
}
\label{fig11}
\end{figure}
Increasing the electron-phonon coupling to the point where 
$\bar{\lambda}>1/\sqrt{K}$ and $g^2>1$, polaron formation starts 
and the LDOS fragments into
an increasing number of polaronic subbands~(Sumi 1974, 
Ciuchi {\it et al.} 1997).
The lowest subbands are either completely or partly 
coherent, i.e. ${\rm Im}\Sigma_i(\omega)=0$, whereas the
higher order subbands are strongly damped because 
${\rm Im}\Sigma_i(\omega)\neq 0$. 
Thus, in the polaron regime, we have two energy 
scales: the
bare bandwidth and the much smaller width of the polaronic subbands. 
Obviously, the width of the distribution of the on-site energies 
$\epsilon_i$ 
can be small on the scale of the bare bandwidth but  
large on the scale of the subband. We distinguish 
therefore between two regimes: The Holstein regime, where 
disorder is small on the scale of the bare bandwidth and
the Anderson regime, where it is large on the scale 
of the width of the subband.  

\subsubsection{Holstein regime}

For our purpose the lowest polaronic subband is of particular 
interest, because it is always completely coherent, i.e. 
${\rm Im}\Sigma_i(\omega)=0$ throughout the subband, and no 
inelastic polaron-phonon scattering interferes with the 
localization properties of the polaron. 

First, we consider the {\it anti-adiabatic strong coupling} regime, 
where the phonon admixture of the states comprising
the lowest polaronic subband is almost energy independent. 
Phonon-induced
long-range tunneling as well as 
band flattening~(Stephan 1996, Fehske and Wellein 1997a)
are absent and the 
LDOS without disorder is rather symmetric. 
Numerical simulations~(Stephan 1996, Fehske and Wellein 1997a)
of the 
Holstein model indicate that the lowest subband in that
case is just a rescaled bare band. Accordingly, we expect
disorder to affect the polaronic subband in the same way as 
it affects the band of the pure Anderson model.
Most notably, the mobility edges should simultaneously appear 
on both sides of the subband. Furthermore, the mobility edge 
trajectories should show the characteristic reentrance behaviour 
at the band edges and the critical disorder 
needed to localize all states of the subband should be 
determined by the states in the centre of the subband. 
In fact, with an appropriate scaling, the mobility edge
trajectories for the lowest polaronic subband and the pure 
Anderson model should coincide. 
We now demonstrate that this is indeed the case.  
\begin{figure}[h]
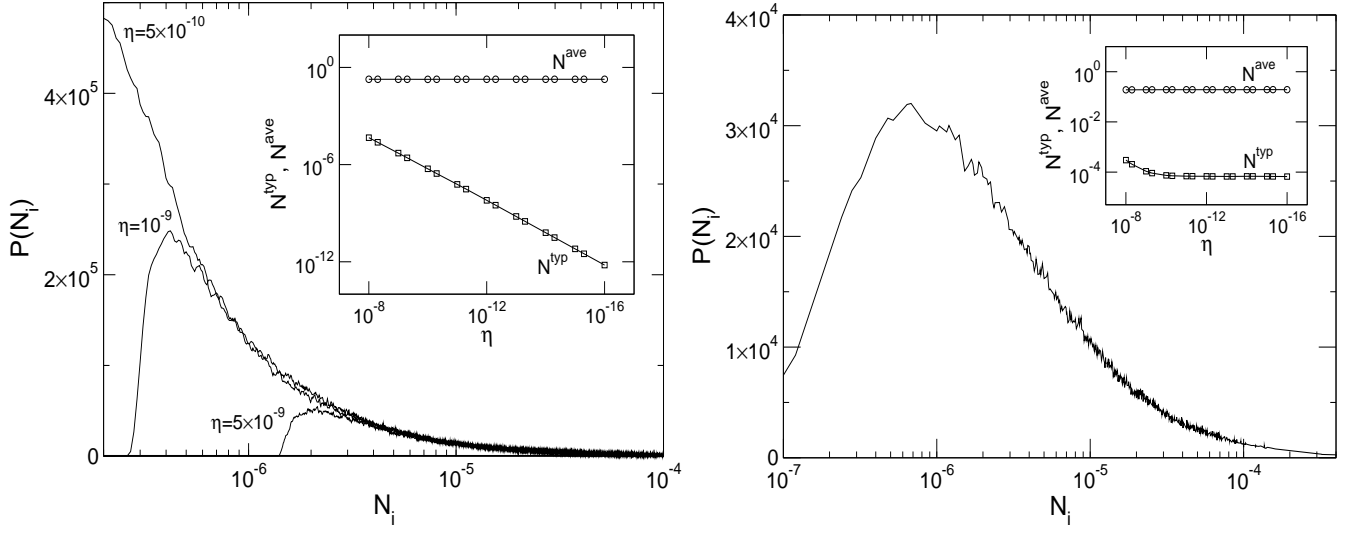

\begin{minipage}{0.5\linewidth}
\psfig{figure=Fig9a.eps,height=7.0cm,width=1.0\linewidth,angle=0}
\end{minipage}
\begin{minipage}{0.48\linewidth}
\psfig{figure=Fig9b.eps,height=7.0cm,width=1.0\linewidth,angle=0}
\end{minipage}
\caption[fig12]
{The left and right panels show, respectively, the
$\eta$-scaling of the distribution for the LDOS at the two
energies indicated in figure~\ref{fig11}.
The polaron parameters are the same as in figure~\ref{fig11}.
In the left panel $\omega=-4.50725$ (above the upper mobility
edge), while in the right panel $\omega=-4.50759$
(below the upper mobility edge).
The insets show the
corresponding average and typical LDOS as a function of $\eta$.}
\label{fig12}
\end{figure}
We start with figure~\ref{fig11}, which shows, for
$\bar{\alpha}=2.25$ and $\bar{\lambda}=9$, the average and typical LDOS for
the lowest subband;  
$\bar{\gamma}=2.5\times W$ with $W=3.45\times 10^{-4}$ the width of 
the subband in units of $W_0=4\bar{J}$. For 
comparison, we also show the DMFT result for the LDOS. As 
expected, the average LDOS is rather symmetric, the small 
asymmetry reflecting the fact that we are not yet in the extreme 
anti-adiabatic regime. As far as the effect of disorder is 
concerned, we see the same overall features as in the 
pure Anderson model (cf. figure~\ref{fig6}): the appearance
of Lifshitz tails and of mobility edges. The typical 
LDOS, shown in figure~\ref{fig11} for 
$\eta=10^{-8}$ and $\eta=10^{-14}$, vanishes on the both sides 
of the subband, clearly indicating the existence of mobility 
edges on both sides of the subband. 
As discussed before, a precise determination of the 
position of the mobility edges requires either to calculate 
the typical LDOS for $\eta=0$ and to track the 
iteration flow with sample generation, or 
to follow the $\eta$-scaling of the LDOS. 
\begin{figure}[h]
\hspace{0.5cm}\psfig{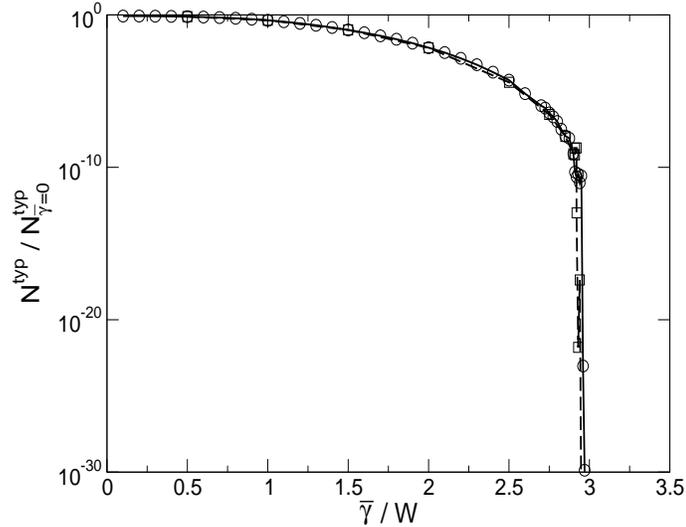}
\caption[fig13]
{The typical LDOS at $\omega=-4.50759$ (roughly the centre of the subband)
scaled to
its value for $\bar{\gamma}=0$ as a function of disorder measured in units of
the width of the subband $W$; $\bar{\alpha}=2.25$ and $\bar{\lambda}=9$.
For comparison, we also plot the scaled typical LDOS for the pure
Anderson model as a function of scaled disorder.
}
\label{fig13}
\end{figure}
For a quantitative calculation of the mobility edges, 
we adopted here the $\eta$-scaling approach. 
In figure~\ref{fig12} we show, for the two representative energies 
indicated in figure~\ref{fig11}, 
the $\eta$-scaling of 
the distribution for the typical LDOS, the typical LDOS, and the 
average LDOS. Above the upper mobility edge at $\omega=-4.50725$, the 
distribution shows the characteristic properties associated with localized 
states: An extremely small most probable value and a long tail 
resulting in an average value much larger than the most 
probable value. The typical LDOS again scales to the fixed $\eta$ 
value, while the average LDOS is independent of $\eta$. For 
the extended states at $\omega=-4.50759$, on the other hand, 
we again see that the 
distribution is insensitive to $\eta$ and both average and 
typical LDOS converge to finite values independent of $\eta$.

As for the pure Anderson model, tracking the 
$\eta$-scaling of the LDOS for all energies comprising the 
subband, enables us to map out the mobility 
edge trajectory for the subband. From physical considerations, we 
know that the trajectory should be the same as for the pure 
Anderson model. Instead of calculating the whole mobility edge
trajectory, it is therefore sufficient to verify, for representative 
energies, that the two mobility edge trajectories indeed coincide.

As a first demonstration of this assertion, we plot 
in figure~\ref{fig13} the typical LDOS for $\omega=-4.50759$ 
(roughly the centre of the subband) scaled to its value at $\bar{\gamma}=0$ as 
a function of disorder in units of the width of the subband and compare it 
with the scaled typical LDOS at $\omega=0$ for the pure Anderson model, 
again as a function of $\bar{\gamma}$ scaled to the bandwidth. 
As anticipated, the two plots coincide. As in the pure Anderson 
model, the state at the centre of the (sub)band is localized for 
$\bar{\gamma}/W\ge 2.9$. 
\begin{figure}[h]
\hspace{0.5cm}\psfig{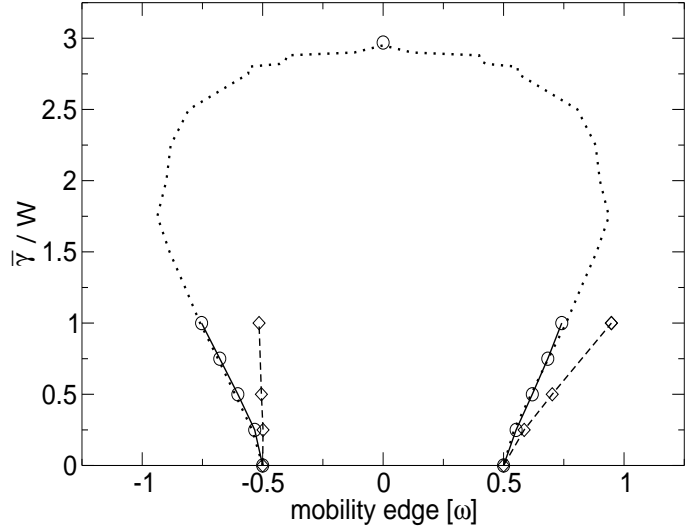}
\caption[fig14]
{Parts of the mobility edge trajectories for the lowest polaronic subband of
the Anderson-Holstein model in the
anti-adiabatic strong coupling ($\bar{\alpha}=2.25$ and $\bar{\lambda}=9$)
and the adiabatic
intermediate-to-strong coupling regime ($\bar{\alpha}=0.2$ and $\bar{\lambda}=1$). Disorder
is measured in units of the respective width of the  subbands, which, for the plot,
are scaled to $W_0=4\bar{J}=1$.
}
\label{fig14}
\end{figure}
To provide further evidence for our assertion that the 
localization properties of the lowest subband in the 
anti-adiabatic strong coupling regime are the ones of a 
rescaled Anderson model, we explicitly calculated for a 
few selected energies the 
critical disorder needed to localize the states at these energies and, 
thereby, constructed parts of the mobility edge trajectory for the 
lowest subband. Note the extremely high precision of our approach,
which enables us to determine the mobility edge 
trajectory of a subband whose width without disorder 
is $3.45\times 10^{-4}$ in units of $W_0=4\bar{J}=1$. 
The results are shown in figure~\ref{fig14} (cf. circles). 
Obviously, the data points for the subband follow 
exactly the trajectory of the pure Anderson model, even at 
very strong disorder, where all states of the 
subband are localized (see data point at $\bar{\gamma}/W\sim 2.9$). 
In the 
anti-adiabatic strong coupling regime, the localization behaviour 
of the states of the lowest subband is therefore 
identical to the behaviour of the states in a rescaled Anderson 
model. The polaron effect, in particular the band collapse, 
only changes the energy scale on which 
localization takes place, the internal structure of the polaron 
states does not affect the localization properties.   
Thus, in the anti-adiabatic strong coupling
regime, disorder can be sufficiently strong 
to localize all states of the lowest subband and yet too small 
to interfere with the internal structure of the polaron.  

This is not the case in the 
{\it adiabatic intermediate-to-strong coupling} regime, where we find 
substantial differences between the localization properties 
of the lowest polaronic subband and the band of the pure
Anderson model. The differences originate in the changing 
composite structure of the polaron states within the 
lowest subband. In particular at the top of the 
subband, where the band flattening due to the
hybridization with the optical phonon branch significantly 
modifies the LDOS, the localization properties deviate 
strongly from a rescaled pure Anderson model. But also
at the bottom of the subband significant deviations 
can be observed because of the phonon-induced 
long-range tunneling processes.  

In the upper panel of figure~\ref{fig15} we show, for $\bar{\alpha}=0.2$, 
$\bar{\lambda}=1$, and $\bar{\gamma}=0.25\times W$, the average and typical 
LDOS for the lowest subband. Here, $W=8.123\times 10^{-3}$ is the 
width of the subband for the pure Holstein model (in units of 
$W_0=4\bar{J}=1$). In contrast to the anti-adiabatic
strong coupling case, the LDOS is strongly asymmetric, 
a direct result of the band flattening which yields a steeple
at the top of the LDOS. States in the steeple,
belonging to the flat part 
of the dispersion, are already sluggish and therefore 
very susceptible to disorder. The
typical LDOS vanishes therefore rapidly at the top of the subband.
At the bottom, however, where states are rather mobile due to 
phonon-induced 
long-range tunneling, and hence less affected by disorder,  
the typical LDOS is finite.
\begin{figure}[t]
\hspace{0.5cm}\psfig{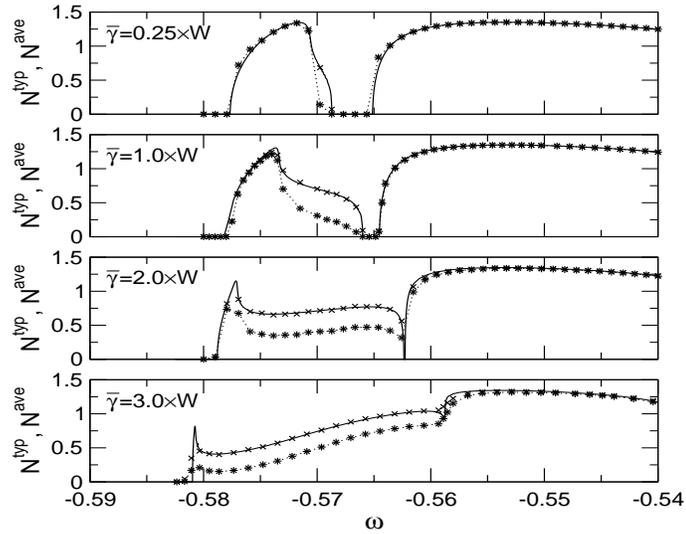}
\caption[fig15]
{Average (crosses) and typical (stars) LDOS for $\bar{\alpha}=0.2$, $\bar{\lambda}=1$ and
four different values
for the disorder strength: $\bar{\gamma}=0.25\times W$,
$\bar{\gamma}=1\times W$, $\bar{\gamma}=2\times W$, and $\bar{\gamma}=3\times W$
(from top to bottom),
with  $W=8.123\times 10^{-3}$
the width of the lowest
polaron subband of the pure Holstein model (in units $W_0=4\bar{J}=1$).
The solid line is the DMFT result for the LCDOS.
}
\label{fig15}
\end{figure}
To determine the position of the mobility edges we
again employ the $\eta$-scaling approach. 
In figure~\ref{fig14} we show for $\bar{\alpha}=0.2$ and $\bar{\lambda}=1$ 
parts of the mobility edge trajectory for the lowest 
subband in the adiabatic strong coupling regime (cf. diamonds). As  
expected, the states at the bottom of the 
subband are almost insensitive to small amounts of disorder, resulting
in a lower mobility edge which is pinned at the lower band edge,
which in fact does also not change. 
States at the top 
of the subband, in contrast, are immediately affected by 
disorder. Small amounts of disorder (small even on the scale of 
the subband) are sufficient to shift the upper band edge to higher 
energies and to localize the states at the top of the subband.
The upper mobility edge moves very fast away 
from the upper band edge. As a result, the mobility edge  
trajectory for the lowest subband in the adiabatic strong coupling 
regime is very asymmetric (for small disorder). 

For large disorder (not shown in figure~\ref{fig14}), 
of the order of the width of the subband, 
the upper band edge of the disorder-broadened
lowest subband moves into a spectral range with significant
inelastic scattering. The reason is the following: Without disorder, 
subbands are 
separated by singularities in the imaginary part of the 
electron-phonon self-energy. Increasing disorder redistributes
states. In particular, it moves states into the gap region,  
which separated the first from the second subband, thereby giving 
rise to a shrinking gap. 
Concomitantly, the singularity in the electron-phonon
self-energy broadens, its state repelling character weakens, 
resulting in an enhanced inelastic scattering rate for the 
states at the top of the lowest subband, which strongly  
suppresses localization in this spectral range.
Both effects, the merging of the subbands and the increased 
inelastic scattering, occur before a reentrance behaviour of the
mobility edge trajectories can be observed. 

To support the scenario just described we depict in the 
three lower panels of figure~\ref{fig15}, 
for polaron parameters $\bar{\alpha}=0.2$ and $\bar{\lambda}=1$, the 
typical and
average LDOS for $\bar{\gamma}=1\times W$, $2\times W$, 
and $3\times W$. Again, $W=8.123\times 10^{-3}$ is the width of the 
subband for the pure Holstein model (in units of 
$W_0=4\bar{J}=1$). In all three 
panels, all states of the lowest subband are now delocalized, the typical 
LDOS is finite for all energies comprising the lowest subband.  
For the upper panel, where a small gap still  
separates the two lowest subbands, the asymmetric localization behaviour 
of the states at the bottom and the top of the lowest subband 
is still visible. Nevertheless, inelastic scattering is so strong  
as to make the typical LDOS finite even at the top of the subband.
With increasing disorder, shown in the three lower panels, the gap
vanishes. More importantly, however, the typical LDOS increases
with increasing disorder. Thus, in this regime, disorder delocalizes 
states. Eventually, of course, with disorder on the scale of the bare 
bandwidth, all states would be localized. However, the critical disorder 
strength
is larger than in the case of the pure Anderson model, because 
the inelastic scattering has to be overcome.
  
The merging of the subbands signals that the internal structure of 
the polaron states and disorder start to strongly interfere 
with each other. With the vanishing subband structure, the concept
of a subband mobility edge trajectory breaks down and a 
reentrance behaviour of the mobility 
edge trajectory cannot be established.  

\subsubsection{Anderson regime}

The vanishing subband structure characterizes a transition regime,
between the Holstein and the Anderson regime, where disorder is
on the scale of the energy gap between the subbands.  
The localization properties in the 
transition region are rather complicated and 
beyond the scope of this paper. The situation
becomes more transparent in what we call the Anderson regime, where 
disorder is large compared to the width of the polaronic 
subbands.
\begin{figure}[h]
\hspace{0.5cm}\psfig{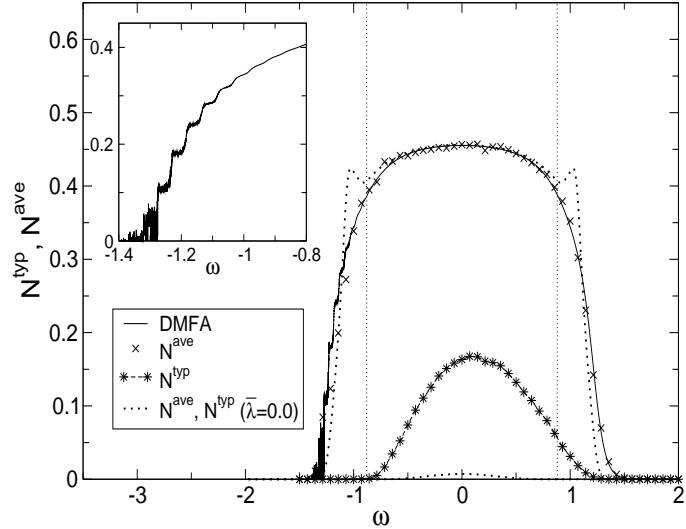}
\caption[fig16]
{Average (crosses) and typical LDOS (stars) in the Anderson regime for
$\bar{\alpha}=0.2$,
$\bar{\lambda}=0.75$, and $\bar{\gamma}=2$. The solid line is the LCDOS obtained from DMFT
(a blow-up of the low-energy part is shown in the inset).
Dashed lines depict the average and typical LDOS for $\bar{\lambda}=0$, respectively, and
the vertical dotted lines indicate the mobility edges of the noninteracting system.
}
\label{fig16}
\end{figure}
In figure~\ref{fig16} we show the typical and average LDOS for 
$\bar{\alpha}=0.2$, $\bar{\lambda}=0.75$, and $\bar{\gamma}=2$. 
Without electron-phonon coupling, 
there would be mobility edges at $\omega\approx\pm 0.9$. In the presence 
of electron-phonon coupling, we first note that the symmetry between the 
lower and upper mobility edges is broken. The lower mobility edge 
is located at 
$\omega\approx -0.8$, which is above the mobility edge of the noninteracting
system. (Here, we did not perform the $\eta$-scaling necessary for a 
precise determination of the mobility edge.)
Thus, at the low energy edge, electron-phonon coupling enhances the
tendency towards localization. The upper mobility edge, in contrast, is shifted 
to higher energies ($\omega\approx 1.2$), i.e., at the upper band edge 
electron-phonon coupling delocalizes states and works against localization.  
If we mapped out a mobility edge trajectory,
we would find a pronounced asymmetry between the lower and the 
upper part of the trajectory.

The asymmetry can be explained, if we recall that our calculation is for 
$T=0$. Therefore, states at energy 
$\omega$ can, due to electron-phonon interaction, only couple to states at 
energies less than $\omega$. This leads at the high energy side of the
LDOS to a phonon-induced coupling of localized states above the  
mobility edge of the noninteracting system to delocalized states below. As a consequence, 
the localized
states become delocalized. 
On the low energy side, the situation
is different. States below the lower mobility edge of the noninteracting system
remain localized, because they can only couple to states which are already
localized. Above the 
lower mobility edge of the noninteracting system, electron-phonon interaction
attempts however to
transform electronic band states into polaronic (sub)band states,
as suggested by Anderson~(1972).
Hence, these states
become heavier and
more susceptible to disorder. As a consequence, the lower mobility
edge of the interacting system shifts
above the lower mobility edge of the noninteracting one.   

Further insight about the nature of the states can be gained from the LCDOS 
obtained within the DMFT (cf. solid line in figure~\ref{fig16} and inset). 
Due to electron-phonon coupling, the DMFT results for the LCDOS show 
at the low energy side pronounced plateaus with a width given by the
phonon energy. (The spikes are numerical artifacts.) The step-like
increase of the DMFT LCDOS, together with the vanishing of the typical LDOS,
is a clear signature for localized polaron states. The  
character of the states is here revealed by the DMFT LCDOS: It arises from 
polaronic defects [cf. independent boson model~(Mahan 1990)], which are centered 
around different on-site energies, because of the disorder. Since they are 
decoupled, the
DMFT LCDOS (with a higher resolution we would see the same in 
the average LDOS) does not change with energy as long
as the fluctuations of the on-site energies are smaller than the phonon
energy. If the difference in on-site energy is equal to the phonon
energy, a step arises because states with one additional phonon contribute. 
The step heights reflect therefore the phonon distribution of the 
polaronic defect states.   

Whereas in the adiabatic intermediate-to-strong coupling regime {\it disorder} 
affects states differently at the bottom and the top of the polaronic subband, 
in the Anderson regime, {\it electron-phonon coupling} affects states in the 
vicinity of the lower and upper mobility edges of the noninteracting system
differently. 
Only in the vicinity of the lower mobility edge of the noninteracting system,
electron-phonon coupling and localization work in the same direction. At the
upper mobility edge, electron-phonon coupling
in fact delocalizes states. 

\section{Conclusions}

We have adopted the statDMFT to the Anderson-Holstein
model to develop a self-consistent theory of 
localization in a generic disordered electron-phonon
system. In particular, we investigated the 
localization properties of a single electron over
a wide range of polaron parameters (adiabaticity 
and electron-phonon coupling strength). 

Our approach is non-perturbative in the electron-phonon coupling
and accounts for the spatial fluctuations of the 
environment by promoting all variables of the 
theory to random variables. The object of the 
theory is to calculate distributions (random samples).
Of particular importance are the distributions for the 
LDOS, LCDOS, and the tunneling rate. Whereas  
information about the spatial dependence of the wave function
$\Psi_i$ can be extracted from the distribution for the LDOS, 
i.e. $\Psi_i\leftrightarrow P(N_i)$, the 
configuration averaged, spectrally resolved
return probability is closely related to the distributions for 
the LCDOS and the tunneling rate, 
i.e. $f_{ii}\leftrightarrow P(N_i^{(j)}), P(\Gamma_i)$
[cf. Eq.~\ref{fii}].

In this paper, we focused on the distribution for 
the LDOS. Since it is a measure of  
the spatial distribution of the wave function, it contains
direct information about the localization
properties: States are localized if the distribution
of the LDOS is very asymmetric with an extremely long tail.
Although the tail, which is due to the few sites on which 
(localized) wave functions are finite, makes the 
average LDOS finite, the more representative typical LDOS
vanishes. Delocalized states, on the other hand, are 
characterized by a symmetric distribution of the LDOS for 
which the typical and average LDOS are of the same order
of magnitude. 

Using the typical LDOS to distinguish localized from 
extended states, we investigated in detail the 
localization properties of a single electron in 
the Anderson-Holstein model. We distinguished two
parameter regimes: The weakly disordered Holstein 
regime, where 
disorder is small on the scale of the bare  
bandwidth and the polaronic subband structure is well 
developed, and the strongly disordered Anderson regime, with 
disorder large on the scale of the width of the subband, strong enough
to interfere with the internal structure of the 
polaron states.  

In the weakly disordered Holstein regime, 
we focused on a comparison of the mobility 
edge trajectories for the lowest polaronic subband and
the bare band of the Anderson model. 
We found significant deviations, most notably in the
adiabatic, intermediate-to-strong coupling regime. 
Two main conclusions can be drawn from our studies:
In the strong coupling, anti-adiabatic regime, the 
internal structure of the polaron is irrelevant and
the localization properties of the lowest polaronic 
subband are essentially the ones of a rescaled 
Anderson model. The overall scale of the disorder
is of course much smaller in the polaron case, 
suggesting that Holstein polarons
are most probably always localized at $T=0$. 
The internal structure of the
phonon dressing contributes
the most in the adiabatic intermediate-to-strong coupling regime.
Initially, for small disorder $\bar{\gamma}<W$ ($W$ the width of the lowest 
subband in the clean system), states at the high-energy edge of the 
lowest polaronic subband are extremely 
sensitive to disorder and rather easy to localize, whereas
states at the bottom are almost insensitive to disorder. In
contrast to the pure Anderson model, the mobility edge 
trajectory (in the disorder range where it can be 
defined) is not symmetric with 
respect to the band centre. Whereas the 
lower mobility edge is pinned to the lower band edge, 
the upper mobility edge moves quickly away 
from the upper band edge. A reentrance of the
upper mobility edge towards the centre of the subband 
cannot be observed, because at larger disorder strength subbands 
vanish and the concept of a subband mobility edge trajectory 
obviously breaks down.  

The merging of polaron subbands is characteristic for a 
transition regime where disorder is of the order of the
energy gap and strongly interferes with the 
internal structure of the polaron states. The detailed   
investigation of this regime is beyond the scope of this paper. 
Instead we presented results for the 
Anderson regime, where disorder is much larger than the 
width of the subbands. In this regime disorder and electron-phonon 
coupling work 
in the same direction at the bottom of the band, where 
electron-phonon coupling even enhances the tendency towards localization, 
and against each other at the top, where electron-phonon coupling 
delocalizes states above the upper mobility edge of the noninteracting
system. In the Anderson regime, 
the mobility edge trajectory would acquire therefore a pronounced
asymmetry. 

Several issues must be however clarified before real materials
with polaronic excitations
can be analysed along the lines presented in 
this paper. In most polaronic 
materials electron densities are finite. The most pressing issue 
is therefore the consideration of the 
polaron-polaron interaction, 
mediated either due to phonons or due to the Coulomb potential
of the electron charge. In principle, the statDMFT is capable 
of tackling this challenging problem. The main obstacle here
is the controlled calculation of the interaction  self-energy
for the ensemble of impurity models.
The exact continued fraction expansion works only for a single
electron and cannot be adopted to finite densities. 
For finite densities, the interaction  self-energy has to be 
obtained by perturbation theory. To overcome the limitations
of the perturbation theory, it is also conceivable  
to use direct numerical simulations of the ensemble of the
local Green function, based, for instance, on
Quantum-Monte-Carlo or Exact Diagonalization techniques.
The localization properties per se can again be simply extracted 
from an analysis of the distribution for the LDOS.  

\section*{Acknowledgments}

We appreciate useful and stimulating discussions with 
Prof. W. Weller. We also thank Prof. A. S. Alexandrov and 
Dr. A. Saxena for 
critically reading the manuscript. One of us (A.A.) acknowledges 
support from the European Graduate School, Bayreuth. 

\appendix 

\section{DMFT limit}

For a lattice with infinite coordination number, the statDMFT
reduces to the DMFT. To demonstrate this point, it is convenient
to scale the transfer amplitude
$J\rightarrow \bar{J}/\sqrt{K}$~(Bronold and Fehske 2002).

Restricting for simplicity the discussion to the Bethe lattice, 
Eq. (\ref{Gii2})
shows that for $K\rightarrow \infty$ cavity and local Green
function are identical and Eq. (\ref{Gii1}) becomes
\begin{eqnarray}
G_{i\sigma i\sigma}(i\omega_n)=
{1\over{i\omega_n+\mu-\epsilon_i-\bar{J}^2G_\sigma^{\rm ave}(i\omega_n)
-\Sigma_\sigma(i\omega_n)}},
\label{DMFT1}
\end{eqnarray}
where we have used the central limit theorem to replace the
hybridization function on the rhs by the average of the local Green
function. At this point the scaling of $J$ is most convenient.                

The self-energy is calculated from the effective single-site
action $S(i)$ where, again due
to the central limit theorem, the hybridization function is replaced by
$\bar{J}G^{\rm ave}_\sigma$. The self-energy is therefore
a functional of $G^{\rm ave}_\sigma(i\omega_n)$, i.e., on the rhs of
Eq. (\ref{DMFT1}) the only random variable is $\epsilon_i$.
The sample average $(1/K)\sum_{j=1}^K (...)$ of Eq. (\ref{DMFT1})
is therefore identical to the site average over $\epsilon_i$.
Denoting the site average by $\langle...\rangle_{\epsilon_i}$, we get
\begin{eqnarray}
G_{\sigma}^{\rm ave}(i\omega_n)=
\bigg\langle\left[
i\omega_n+\mu-\epsilon_i-\bar{J}^2G_\sigma^{\rm ave}(i\omega_n)
-\Sigma_\sigma[\epsilon_i,G_\sigma^{\rm ave}]
\right]^{-1}\bigg\rangle_{\epsilon_i},
\label{DMFT2}
\end{eqnarray}
which, together with
\begin{eqnarray}
S(i)\bigg|_{K\rightarrow\infty}&=&\int_0^\beta d\tau d\tau'\sum_\sigma
c_{i\sigma}^\dagger(\tau)\bigg\{
[\partial_\tau+\epsilon_i-\mu]\delta(\tau-\tau')
+\bar{J}^2G_\sigma^{\rm ave}(\tau-\tau')\bigg\}c_{i\sigma}(\tau')
\nonumber\\
&+&S_{int}(i)
\label{DMFT3}
\end{eqnarray}
constitutes the DMFT equations for the average local Green
function $G_{\sigma}^{\rm ave}(i\omega)$. 
For a single electron, Eqs. (\ref{DMFT2})
and (\ref{DMFT3}) reduce, for the case without disorder, to 
the dynamical coherent-potential approximation (DCPA) equations~(Sumi 1974)
and, for the case without electron-phonon coupling, to 
the coherent-potential approximation (CPA) equations~(Elliott {\it et al.} 1974).
\begin{figure}[t]
\hspace{0.5cm}\psfig{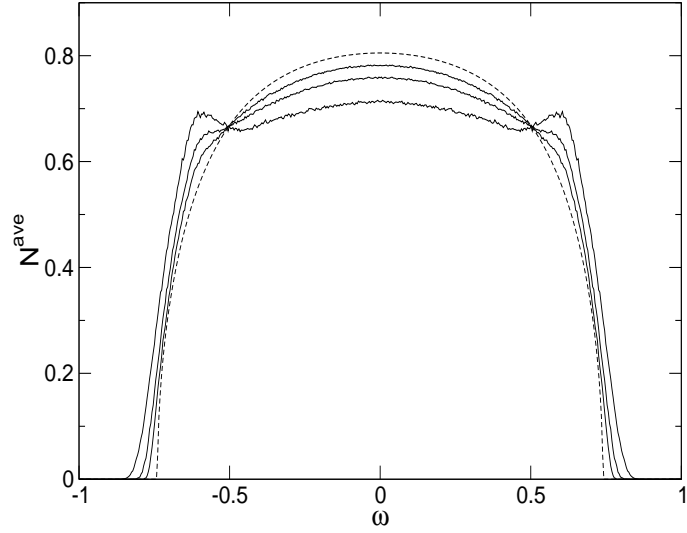}
\caption[figAppB]
{Average LDOS for the Bethe lattice with connectivity
$K=2, 4$ and $8$ (solid lines form top to bottom).
The dashed line is the DMFT
LDOS, i.e. the LDOS for $K\rightarrow\infty$.}
\label{figAppB}
\end{figure}   
For the pure Anderson model this is illustrated in figure~\ref{figAppB}.
With increasing lattice coordination number $K$ we observe two effects.
First, the difference between cavity and local Green function
indeed disappears, as can be seen from the vanishing of
the bumps due to the van-Hove singularities; second, the local density of states
converges to the CPA density of states.  Note also, Lifshitz
tails, which are present within the statDMFT,
also vanish, as expected, with increasing $K$.  

\newpage


\section{References}

\vspace{0.5cm}
\noindent
Abrahams, E., Anderson, P. W., Licciardello, D. C., and
Ramakrishnan, T. V., 1979, Phys. Rev. Lett. {\bf 42}, 673.

\vspace{0.5cm}
\noindent
Abou-Chacra, R.,
Anderson, P. W., and Thouless, D. J.,
1973, J. Phys. C {\bf 6}, 1734.

\vspace{0.5cm}
\noindent
Aguiar, M. C. O., Miranda, E., and Dobrosavljevi\'c, V., 2003, 
Phys. Rev. B {\bf 68}, 125104. 

\vspace{0.5cm}
\noindent
Alexandrov, A. S., Kabanov, V. V., and Ray, D. K., 1994,
Phys. Rev. B {\bf 49}, 9915.

\vspace{0.5cm}
\noindent
Alexandrov, A. S., and Kornilovitch, P. E., 1999, Phys. Rev. Lett. {\bf 82},
807.

\vspace{0.5cm}
\noindent
Alvermann, A., 2003, Diplom thesis, Universit\"at Bayreuth,
unpublished.

\vspace{0.5cm}
\noindent
Anderson, P. W., 1958, Phys. Rev. {\bf 109}, 1498.

\vspace{0.5cm}
\noindent
Anderson, P. W., 1972, Nature (London) Phys. Science {\bf 235},
163.          

\vspace{0.5cm}
\noindent
Bar-Yam, Y., Egami, T., Mustre-de~Leon, J., and Bishop, A. R. (editors),
1992, {\it Lattice Effects in High-$T_c$ Superconductors}
(Singapore: World Scientific). pp. 377-422, and references therein.

\vspace{0.5cm}
\noindent
Belitz, D., and Kirkpatrick, T. R., 1994,
Rev. Mod. Phys. {\bf 66}, 261.

\vspace{0.5cm}
\noindent
Bonc\u{a}, J., Trugman, S. A., and Batisti\'{c}, I., 1999,
Phys. Rev. B {\bf 60}, 1633.

\vspace{0.5cm}
\noindent
Brezini, A., 1982, J. Phys. C {\bf 15}, L211.

\vspace{0.5cm}
\noindent
Brezini, A., and Olivier, G., 1983, Phil. Mag. B {\bf 47}, 461.

\vspace{0.5cm}
\noindent
Brezini, A., and Zerki, N., 1992, phys. stat. sol. (b) {\bf 169}, 253.      

\vspace{0.5cm}
\noindent
Bronold, F. X.,
Saxena, A., and Bishop, A. R.,
2001, Phys. Rev. B {\bf 63}, 235109.

\vspace{0.5cm}
\noindent
Bronold, F. X.,  and Fehske, H., 2002,
Phys. Rev. B {\bf 66}, 073102; {\it ibid.}, 2003,
Acta Phys. Pol. {\bf 34}, 851.

\vspace{0.5cm}
\noindent
Capone, M., Stephan, W., and Grilli, M., 1997, Phys. Rev. B {\bf 56}, 4484.

\vspace{0.5cm}
\noindent    
Castellani, C., Di
Castro, C., Lee, P. A., and Ma, M., 1984, Phys. Rev. B {\bf 30}, 527.  

\vspace{0.5cm}
\noindent
Cheong, S.-W., Hwang, H. Y., Chen, C. H., Batlogg, B., Rupp Jr., L. W.,
and Carter, S. A., 1994, Phys. Rev. B {\bf 49}, 7088.

\vspace{0.5cm}
\noindent
Ciuchi, S., de Pasquale, F., Fratini, S., and Feinberg, D., 1997,
Phys. Rev.
{\bf B 56}, 4494.

\vspace{0.5cm}
\noindent
Cohen, M. H., Economou, E. N., and
Soukoulis, C. M., 1983, Phys. Rev. Lett. {\bf 51}, 1202.           

\vspace{0.5cm}
\noindent
DeRaedt, H., and Lagendijk, A., 1983, Phys. Rev. B {\bf 27}, 6097.

\vspace{0.5cm}
\noindent
Dobrosavljevi\'c, V.,  and Kotliar, G., 1997, Phys. Rev. Lett. {\bf 78},
3943; {\it ibid.}, 1998, Philos. Trans. R. Soc. Lond. Ser. A {\bf 356}, 1.

\vspace{0.5cm}
\noindent
Economou, E. N., and Cohen, M. H., 1972, Phys. Rev. B {\bf 5}, 2931.

\vspace{0.5cm}
\noindent
Elliott, R. J., Krumhansl, J. A., and Leath, P. L., 1974,
Rev. Mod. Phys. {\bf 46}, 465.

\vspace{0.5cm}
\noindent
Elyutin, P. V., 1979, Sov. Phys. Solid State {\bf 21}, 1590.
{\it ibid.}, 1981, J. Phys. C {\bf 14}, 1435.

\vspace{0.5cm}
\noindent
Fehske, H., Loos, J., and Wellein, G., 1997a, Z. Phys. B
{\bf 104}, 619; {\it ibid.}, 1997b,
Phys. Rev. B {\bf 56}, 4513.

\vspace{0.5cm}
\noindent
Fehske, H., Loos, J., and Wellein, G., 2000, Phys. Rev. B
{\bf 61}, 8016.                  

\vspace{0.5cm}
\noindent
Finkel'stein, A. M., 1983, Zh. Eksp. Teor. Fiz. {\bf 84}, 168
[Sov. Phys. JETP {\bf 57}, 97 (1983)]

\vspace{0.5cm}
\noindent
Firsov, Y. A., 1975, {\it Polarons} (Moscow: Nauka).

\vspace{0.5cm}
\noindent
Fleishman, L., and Stein, D. L., 1979, J. Phys. C {\bf 12}, 4817.

\vspace{0.5cm}
\noindent
Georges, A., Kotliar, G., Krauth, W., and Rozenberg, M. J., 1996,
Rev. Mod. Phys.
{\bf 68}, 13.

\vspace{0.5cm}
\noindent
Girvin, S. M., and Jonson, M., 1980, Phys. Rev. B {\bf 22}, 3583.

\vspace{0.5cm}
\noindent
Heinrichs, J., 1977, Phys. Rev. B {\bf 16}, 4365.

\vspace{0.5cm}
\noindent
Holstein, T., 1959, Ann. Phys. (N.Y.) {\bf 8}, 325; ibid. 343.             

\vspace{0.5cm}
\noindent
Jin, S., Tiefel, T.~H, McCormack, M.,
Fastnach, R. A., Ramesh, R., and Chen, L. H., 1994, Science {\bf 264},
413.

\vspace{0.5cm}
\noindent
Kamenev, A., and Andreev, A.,
1999, Phys. Rev. B {\bf 60}, 2218.

\vspace{0.5cm}
\noindent
Kirkpatrick, T. R., and Belitz, D., 1990, J. Phys. Cond. Mat. {\bf 2},
5259.

\vspace{0.5cm}
\noindent
Kopidakis, G., Soukoulis, C. M., and Economou, E. N., 1996,
Europhysics Lett. {\bf 33}, 459.

\vspace{0.5cm}
\noindent
Kramer, B., and MacKinnon, A., 1993, Rep. Prog. Phys. {\bf 56}, 1469.

\vspace{0.5cm}
\noindent
Kumar, N., Heinrichs, J., and Kumar, A. A., 1975,
Sol. Stat. Commun. {\bf 17}, 541.

\vspace{0.5cm}
\noindent
L. D. Landau, 1933, Z. Phys. {\bf 3}, 664.         

\vspace{0.5cm}
\noindent
Lee, P. A., and Ramakrishnan, T. V., 1985, Rev. Mod. Phys.
{\bf 57}, 287.

\vspace{0.5cm}
\noindent
Logan, D. E., and Wolynes, P. G., 1985, Phys. Rev. B {\bf 31}, 2437;
{\it ibid.}, 1986,
J. Chem. Phys. {\bf 85}, 937;
{\it ibid.}, 1987, Phys. Rev. B {\bf 36}, 4135.

\vspace{0.5cm}
\noindent
Mahan, G., 1990, {\it Many-Particle Physics} (New York: Plenum Press).

\vspace{0.5cm}
\noindent
Marsiglio, F., 1993, Phys. Lett. A {\bf 180}, 280.

\vspace{0.5cm}
\noindent
Miller, J. D., and Derrida, B., 1993, J. Stat. Phys. {\bf 75}, 357.

\vspace{0.5cm}
\noindent
Miranda, E., and Dobrosavljevi\'c, V., 2001a, Phys. Rev. Lett. {\bf 86}, 264;
{\it ibid.}, 2001b, J. Magn. Magn. Mat. {\bf 226-230},
110.

\vspace{0.5cm}
\noindent
Mirlin, A., and Fyodorov, Yan. V., 1994, J. Phys. I France {\bf 4}, 655.

\vspace{0.5cm}
\noindent
Montroll, E. W., and Shlesinger, M. F., 1983, J. Stat. Phys. {\bf 32}, 209.

\vspace{0.5cm}
\noindent
Mott, N. F., 1968a, J. Non-Cryst. Solids {\bf 1}, 1;
{\it ibid.}, 1968b, Phil. Mag. {\bf 22}, 7;
{\it ibid.}, 1976, Commun. Phys. {\bf 1}, 203;
{\it ibid.}, 1981, Phil. Mag. B {\bf 44}, 265.

\vspace{0.5cm}
\noindent
Mueller, H., and Thomas, P., 1983, Phys. Rev. Lett.
{\bf 51}, 702.

\vspace{0.5cm}
\noindent
Oyanagi, H., and Bianconi, A. (eds.), 2001,
{\it Physics in local lattice
distortions: Fundamentals and Novel Concepts},
AIP Conference Proceeding {\bf 554}
(American Institute of Physics).

\vspace{0.5cm}
\noindent
Parris, P. E., Kenkre, V. M., Dunlap, D. H.,
2001, Phys. Rev. Lett. {\bf 87}, 126601.             

\vspace{0.5cm}
\noindent
Ranninger, J., and Thibblin, U., 1992, Phys. Rev. B {\bf 45}, 7730.

\vspace{0.5cm}
\noindent
Rashba, E. I., 1982, {\it Excitons}, edited by Rashba, E. I., and
Sturge, D. M.,
(Amsterdam: North-Holland), p. 542.

\vspace{0.5cm}
\noindent
Salje, E. K. H., Alexandrov, A. S., and Liang, W. Y., 1995,
{\it Polarons and Bipolarons in High-$T_c$ Superconductors and
Related Materials} (Cambrigde: Cambridge University Press).

\vspace{0.5cm}
\noindent
Stephan, W., 1996, Phys. Rev. B {\bf 54}, 8981.

\vspace{0.5cm}
\noindent
Sumi, H., 1974, J. Phys. Soc. Jpn. {\bf 36}, 770.

\vspace{0.5cm}
\noindent
Thouless, D. J., 1970, J. Phys. C {\bf 3}, 1559.

\vspace{0.5cm}
\noindent
Thouless, D. J., 1974, Phys. Rep. {\bf 13}, 93.                

\vspace{0.5cm}
\noindent    
Tokura, Y., and Tomioka, Y., 1999, J. Magn. Magn. Mater.
{\bf 200}, 1. 

\vspace{0.5cm}
\noindent
Vollhardt, D.,  and W\"olfle, P., 1980,
Phys. Rev. B {\bf 22}, 4666.

\vspace{0.5cm}
\noindent
Vollhardt, D., and W\"olfle, P., 1992, {\it Electronic Phase
Transitions}, edited by W. Hanke and Yu. V. Kopaev
(Amsterdam: North Holland), chapter 1, pp. 1--78.

\vspace{0.5cm}
\noindent
von Oppen, F., Wettig, T., and M\"uller, J., 1996,
Phys. Rev. Lett.
{\bf 76}, 491.

\vspace{0.5cm}
\noindent 
Wegner, F. J., 1976, Z. Phys. B {\bf 25},
327.   

\vspace{0.5cm}
\noindent
Wellein, G., R\"oder, H., and Fehske, H.,
1996, Phys. Rev. B {\bf 52}, 9666.

\vspace{0.5cm}
\noindent
Wellein, G., and Fehske, H., 1998,
Phys. Rev. B {\bf 58}, 6208.                          

\end{document}